\documentclass[12pt]{article}
\usepackage{amsmath}
\usepackage{amsfonts}
\usepackage{amscd}
\usepackage{amsthm}
\usepackage{amssymb}
\usepackage{latexsym}
\usepackage{graphicx}
\usepackage{graphics}
\usepackage{subeqnarray}
\usepackage[usenames]{color}
\usepackage{colortbl}
\newcommand{\nc}{\newcommand*}

\newtheorem{lemma}{Lemma}[section]
\numberwithin{equation}{section}
\usepackage[cp1251]{inputenc}
\inputencoding{cp1251}
\usepackage[T2B]{fontenc}
\usepackage[russian,english]{babel}
\captionsrussian
\nc{\beq}{\begin{equation}}
\nc{\eeq}{\end{equation}}
\nc{\lb}[1]{\label{#1}}
\nc{\nn}{\nonumber}
\nc{\ts}{\textstyle}
\nc{\ds}{\displaystyle}
\def\suchthat{\stackrel{\circ}{\scriptstyle\circ}}
\nc{\lar}{\leftarrow}
\nc{\rar}{\rightarrow}
\nc{\Rar}{\Rightarrow}
\nc{\half}{\frac{1}{2}}

\def\H{\mathcal{H}}
\def\He{\mathcal{H}_{even}}
\def\Ho{\mathcal{H}_{odd}}
\def\Hco{\mathcal{H}_{\bullet,\,0}}
\def\Hoc{\mathcal{H}_{0,\,\bullet}}
\def\L{\mathrm{L}}
\def\d{\mathrm{d}}
\def\z{\overline{z}}
\def\Z{\overline{Z}}
\def\AG{\mathfrak{A}}

\def\AGt{\widetilde{\mathfrak{A}}_r}
\def\AGts{\widetilde{\mathfrak{A}}_s}

\def\AGte{\widetilde{\mathfrak{A}}_r^{(even)}}
\def\AGto{\widetilde{\mathfrak{A}}_r^{(odd)}}
\def\AGh{\widehat{\mathfrak{A}}_r}
\def\sdsum{\subset\!\!\!\!\!\!+} %semi direct sum
\def\asp{a^+_{sect}}
\def\asm{a^-_{sect}}
\def\aspm{a^{\pm}_{sect}}
\def\asmp{a^{\mp}_{sect}}
\def\arp{a^+_{rad}}
\def\arm{a^-_{rad}}
\def\arpm{a^{\pm}_{rad}}
\def\armp{a^{\mp}_{rad}}
\def\acop{a^+_{\bullet,\,0}}
\def\acom{a^-_{\bullet,\,0}}
\def\acopm{a^{\pm}_{\bullet,\,0}}
\def\aocp{a^+_{0,\,\bullet}}
\def\aocm{a^-_{0,\,\bullet}}
\def\aocpm{a^{\pm}_{0,\,\bullet}}
\def\Poco{P^{(0)}_{\bullet,\,0}}
\def\Pooc{P^{(0)}_{0,\,\bullet}}
\def\lz{\text{\Large$\mathbf{\zeta}$}}
\def\lzz{\text{\Large$\overline{\mathbf{\zeta}}$}}
\nc\Pnck[2]{P^{(#1)}_{\bullet,\,#2}}
\nc\Pnkc[2]{P^{(#1)}_{#2,\,\bullet}}
\nc\Pnkm[3]{P^{#1}_{#2,\,#3}}
\nc\hPnkm[3]{\widehat{P}^{#1}_{#2,\,#3}}

\def\bask{\left\{U_{k,\,0}\right\}_{k=0}^{\infty}}
\def\basl{\left\{U_{0,\,l}\right\}_{l=0}^{\infty}}
\nc{\reff}[1]{(\ref{#1})} % Put parentheses around equation references
\begin{document}
{}
\bigskip \bigskip
\centerline{\Large\bf  V.V.Borzov$^1$, E.V.Damaskinsky$^2$}
\bigskip
\begin{center}
$^1$Department of Mathematics, St.Petersburg University of Telecommunications, 191065, Moika  61, St.Petersburg, Russia; borzov.vadim@yandex.ru

$^2$Department of Natural Sciences, Institute of Defense Technical Engineering (VITI), 191123, Zacharievskaya 22, St.Petersburg, Russia; evd@pdmi.ras.ru
\end{center}
\bigskip\bigskip

\centerline {\Large\bf The algebra of two dimensional generalized}
\bigskip
\centerline {\Large\bf Chebyshev - Koornwinder oscillator
\footnote {This work was done under the partial support of the RFBR grant
12-01-00207а}}
\bigskip

\begin{abstract}
In the previous works \cite{N46,N47}  authors have defined the oscillator-like system that
associated with the two variable Chebyshev-Koornwinder polynomials.
We call this system the generalized Chebyshev - Koornwinder oscillator.
In this paper we study the properties of infinite-dimensional Lie algebra
that is analogous to the Heisenberg algebra for the Chebyshev - Koornwinder oscillator.
We construct the exact irreducible representation of this algebra in a Hilbert space
$\mathcal{H}$ of functions that are defined
on a region which bounded by the Steiner hypocycloid. The functions are square-integrable with respect to
the orthogonality measure for  the Chebyshev - Koornwinder polynomials and
these polynomials form an orthonormalized basis in the space
$\mathcal{H}$. The generalized oscillator which is studied in the work
can be considered as the simplest nontrivial example of multiboson quantum system that is
composed of three interacting oscillators.
\end{abstract}
\newpage

\section{INTRODUCTION}
\label{Intro}

The notion of the quantum harmonic oscillator is one of the cornerstones of
quantum physics. With this concept closely linked:

1) The Hilbert space $\mathcal{H}$ of  oscillator states; the classical Hermite
polynomials, which are orthogonal with respect to the Gaussian measure on the real axis
and forms
the basis in the space  $\mathcal{H}$; the fundamental quantum-mechanical operators
in  $\mathcal{H}$ --- the operators of the coordinate, momentum, and quadratic Hamiltonian.

2)The Fock space and main operators in this space --- ladder operators, the operator of the
number of particles, and identical  operator, which are the generators of the
oscillator algebra (algebra of dynamical symmetries for the quantum system).

Development of quantum physics at the end of the last century, in particular, the emergence of
quantum algebras \cite{N01}-\cite{N04}, has resulted in various generalizations
of the quantum harmonic oscillator. The first meaningful generalization
was the  $q$-oscillator \cite{N05}-\cite{N09} connected with $q$-deformation of
canonical commutation relations of the
harmonic-oscillator algebra. The next step in
generalization the quantum-mechanical oscillator was the construction of
a ladder operators that satisfy certain commutation relations
(i.e. construction of an oscillator-like algebra that generalizes the Heisenberg algebra
\cite{N10}-\cite{N12}) connected with some polynomials from the Askey - Wilson
scheme \cite{N13}-\cite{N15}.
Analysis of the oscillator algebras related to some known (mostly classical)
orthogonal polynomials allowed us to develop a general scheme for construction of the generalized
oscillator, i.e. oscillator-like algebra which is a
Heisenberg algebra generalization  that connected with the given system of
orthogonal polynomials on the real axis \cite{N16}.

In last time increased interest to different applications of orthogonal polynomials
in several variables. For general results concerning such polynomials we refer the reader
to the monographs \cite{N17,N18}.
We note  also recent works \cite{N19,N20}, as well as earlier work on two-variable
Krall - Schaefer polynomials \cite{N21} that associated with integrable systems on
the spaces of constant curvature (see also \cite{N22,N23}).
In this regard, it is natural to extend the construction of the generalized oscillator
connected with orthogonal polynomials in one variable
to the case of polynomials of several variables.
We start by considering a simple but
non-trivial generalizations of the classical polynomials in one variable
to the case of several variables. A wide class of such
polynomials is connected with the root systems of
Lie algebras \cite{N24}-\cite{N27}.

One of the first works in this direction was the  T. Koornwinder article \cite{N28},
the main results of which were presented in \cite{N17}
(see additional details in \cite{N29}-\cite{N31}). In this work \cite{N28}
Koornwinder introduces  orthogonal polynomials that are a natural generalization of the
classical Chebyshev polynomials associated with the
root system of the Lie algebra $sl(3)$.
These polynomials are called
the Chebyshev - Koornwinder polynomials in the following.
In \cite{N27} and \cite{N31}-\cite{N33} these ideas were extended to the case of several variable
analogues of the other classical polynomials (see also \cite{N34}). More results can be found
in the \cite{N35,N37}. We note also the works  \cite{N39}-\cite{N45} related to
the other generalizations of Chebyshev polynomials on the case of several variables.

The purpose of our work consists  in the determination of the algebra of the generalized
Chebyshev - Koornwinder oscillator and study some properties of this algebra.
Considered  oscillator-like system, associated with Chebyshev - Koornwinder polynomials
(hereinafter ChK-polynomials) \cite{N28}, was defined in the our works
\cite{N46,N47}, based on the scheme proposed in \cite{N16}.
The Chebyshev-Koornwinder oscillator (hereinafter ChK-oscillator)
was considered in \cite{N46} as a union of three oscillators: sectorial, radial
and boundary ones. In  \cite{N46} the main attention was paid to the quantum-mechanical
aspects of the generalized oscillator while the details of the construction,
exact formulas for the ladder operators, and the study of related oscillator
algebras were omitted due to space restriction.
In \cite{N47} were obtained differential expression for the ladder operators
of the ChK-oscillator and extended the Koornwinder's
algebra of differential operators \cite{N28}  to Abelian subalgebra of the
ChK-oscillator algebra.
In this work we give a complete description of the
algebras of sectorial, radial, and boundary oscillators, as well as the
construction and study of  algebra of two-dimensional ChK-oscillator.

The article has the following structure.
In Sec.2, we briefly recall some information on  ChK-polynomials of the 2nd type
together with some details of construction of the oscillator algebras necessary
for us in the following.
Below are a few comments regarding the content of Sec.2.

Firstly, we note that the choice of the "coordinate operators" $Z$ and $\overline{Z}$
for sectorial oscillator in Sec.2.2.1 (see formulas \reff{13-2.1}-\reff{13-2.2})
with the help of recurrence relations \reff{13-1.1} is a natural
generalization of the corresponding formula for a one-dimensional coordinate operator
(see \cite{N14,N14a,N16}).

Construction of the corresponding "momentum operators" $P_Z=Z^\dag$ and
$P_{\overline{Z}}=\overline{Z}^\dag$ is performed with the help of the
Poisson kernel \reff{13-2.4}
for the ChK-polynomials and used the method developed in \cite{N16}.
Ladder operators and quadratic Hamiltonian are constructed for operators
$Z, \overline{Z}, Z^\dag, \overline{Z}^\dag$ in the standard way.
To save space, we omit the unused in this work
generalization of the Fourier transform (similar to the transformation
in the work \cite{N16}), based on the use of the Poisson kernel.
Next, note that the correct choice of "coordinate" operator $X$ for radial oscillator
in Sec.2.2.2 (see \reff{13-2.18}) is not as easy as for sectorial oscillator.
The mentioned choice demands prior to obtain the recurrence relations for
the ChK-polynomials in the "radial direction". The momentum operator $X^\dag$
is constructed using the corresponding Poisson kernel \reff{13-2.19}. This results
in the non-standard form for the ladder operators \reff{13-2.23}
and Hamiltonian \reff{13-2.28} of the radial oscillator. The additional term in
\reff{13-2.28} can be considered as "interaction energy" of the sectorial and
radial oscillators.

Secondly, the introduction in the Subsection 2.2.3 one more "marginal" oscillator
due to the fact that according to the \reff{13-1.2} the union $\mathfrak{A}_{s,r}$
of the  sectorial and radial oscillator algebras decomposes into the direct sum
$\mathfrak{A}_{s,r}=\mathfrak{A}^{even}_{s,r}\oplus\mathfrak{A}^{odd}_{s,r}$.
Thus it is clear that $\mathfrak{A}_{s,r}$ is a subalgebra of the desired algebra
of two-dimensional ChK-oscillator. So that we must introduce additional oscillator
in Subsection 2.2.3, which we call the "boundary" oscillator, because the generators
of the corresponding oscillator algebra are not trivial only in "boundary subspaces"
$\Hco$ and $\Hoc$ (see \reff{13-2.31}).

Further, we note that all constructed in section 2.2 algebras of sectorial,
radial and boundary oscillators are associative algebras, which can be
(minimally) extended to some (infinite-dimensional) Lie algebras.
The construction of such extensions requires the introduction of additional
generators satisfying the new commutation relations induced by commutation
relations of considered associative algebras. These constructions are the
main content of Sec.3---Sec.5.

Finally, in section 6, is determined  the Lie algebra of two-dimensional
ChK-oscillator. Note that it is more natural to understood as oscillator algebra not the associative algebra but its minimal extension to the Lie algebra.
We suppose that this is correct because the considered algebra should be a
generalization of the Heisenberg-Lie algebra for
the quantum-mechanical oscillator.
Some properties of this infinite-dimensional Lie algebra are investigated in Sec.7.

\section{Preliminaries}
For the reader convenience, we present briefly some information
(necessary for further discussions ) from the works \cite{N46,N47}.

\subsection{The Chebyshev --- Koornwinder polynomials (ChK-polynomials)}
The Chebyshev --- Koornwinder polynomials (ChK-polynomials) \cite{N28}
can be defined by recurrent relations
\begin{align}\label{13-1.1}\index{1}
z\,U_{k,\,l}(z,\,\z)=U_{k+1,\,l}(z,\,\z)+U_{k,\,l-1}(z,\,\z)+U_{k-1,\,l+1}(z,\,\z),\nn\\
\z\,U_{k,\,l}(z,\,\z)=U_{k,\,l+1}(z,\,\z)+U_{k-1,\,l}(z,\,\z)+U_{k+1,\,l-1}(z,\,\z),
\end{align}
subject to the following conditions
$$
U_{0,\,-1}\!=\!U_{-1,\,0}\!=\!0,\,U_{0,\,0}\!=\!1,\quad
\overline{U_{k,\,l}}(z,\,\z)\!=\!U_{l,\,k}(z,\,\z)\!=\!U_{k,\,l}(\z,\,z).
$$
It is known \cite{N28} that ChK-polynomials form an orthonormal basis in the Hilbert space
$\H=\L^2(S;\mu(\d x\d y)),$
where $S$ --- the region within the Steiner hypocycloid (see Fig.1),
and $\mu$ --- probability measure on $S$,
given by the formula
$$
\mu(\d x,\d y)=\frac{1}{2\pi^2}\sqrt{27-18z\z+4z^3+4\z^3-\z^2z^2}dxdy;\quad (z=x+iy).
$$
The normalization constant in this formula was calculated in \cite{N47} (see also \cite{N45,N24}).
\begin{figure}[h]
\centering
\includegraphics[width=6.0cm]{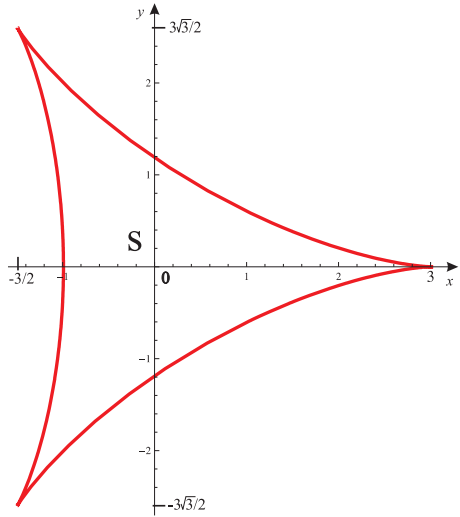}
\caption{Steiner hypocycloid }
\end{figure}

\subsection{Chebyshev --- Koornwinder oscillator (ChK-oscillator)}
In the work \cite{N46} we applied the construction \cite{N16} of the generalized oscillator
connected with a system of orthogonal polynomials on the real axis to the case of
ChK-polynomials of two variables.
Chebyshev --- Koornwinder oscillator (ChK-oscillator)
was considered in \cite{N46} as a union of three oscillators: sectorial,
radial, and boundary ones. To describe them it is convenient to
use  the following decompositions of the space $\H$ (which we consider as the Fock space),
in the direct sums of subspaces:
\begin{equation}\label{13-1.2}
\H=\bigoplus_{N=0}^{\infty}\H^{(N)},\quad \H=\He\bigoplus\Ho,
\end{equation}
where $N$-particle sector $\H^{(N)}$ is the closure in $\H$ of the
linear span of set of basis vectors
 $\left\{U_{k,\,l}(z,\overline{z})\left|\right.k+l=N\right\}$, and subspaces $\He$ и $\Ho$
 are the closures of linear spans of the following sets
\begin{equation*}
\left\{U_{k,\,l}(z,\overline{z})\left|\right.k+l=2n\right\}_{n=0}^{\infty}
\quad \text{and} \quad
\left\{U_{k,\,l}(z,\overline{z})\left|\right.k+l=2n+1\right\}_{n=0}^{\infty},
\end{equation*}
respectively. Moreover, we use the following subspaces
  $\Hco$ and $\Hoc$  which are the closures of linear spans of the sets
\begin{equation*}
\left\{U_{k,\,0}(z,\overline{z})\right\}_{k=0}^{\infty}
\quad \text{and} \quad
\left\{U_{0,\,l}(z,\overline{z})\right\}_{l=0}^{\infty},
\end{equation*}
and the spaces $\mathcal{H}^{\bullet,\,0}$
and $\mathcal{H}^{0,\,\bullet}$ defined by
$$
\mathcal{H}^{\bullet,\,0}=\H\ominus\Hco,\qquad \mathcal{H}^{0,\,\bullet}=\H\ominus\Hoc.
$$
As in the case of the standard quantum-mechanical oscillator,
ChK-oscillator can be determined using the ladder operators.
The ladder operators $\aspm$ of sectorial oscillator leave invariant the subspaces
 $\H^{(N)}$, and the ladder operators
 $\arpm$ of radial oscillator transform the subspaces $\H^{(N)}$ into the
 subspaces $\H^{(N\pm 2)}$. In addition, operators   $\aspm$ and  $\arpm$
leave invariant subspaces $\He$ and $\Ho$. Finally, the ladder operators
$\acopm$ of boundary oscillator transform the subspace $\H^{(N)}\cap\Hco$
in the subspace$\H^{(N\pm 1)}\cap\Hco$ and equal to zero on $\mathcal{H}^{\bullet,\,0}$,
and the ladder operators $\aocpm$ transform the subspace $\H^{(N)}\cap\Hoc$ in the
subspace $\H^{(N\pm 1)}\cap\Hoc$ and equal to zero on $\mathcal{H}^{0,\,\bullet}$.
Action of these operators is schematically shown in figure 2, where the basic
elements $\left\{U_{k,\,l}(z,\overline{z})\right\}_{k,\,l}^{\infty}$ are represented
by points of the rectangular lattice.

\begin{figure}[h]
\centering
\includegraphics[width=8.0cm]{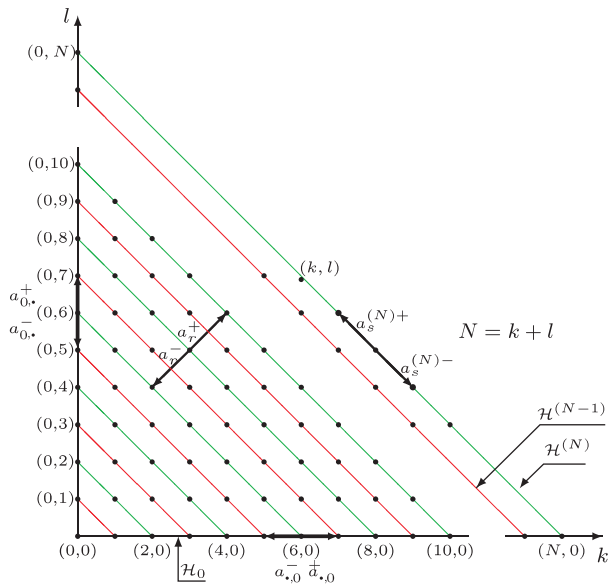}
\caption{Action of ladder operators}
\end{figure}

\subsubsection{Generalized sectorial oscillator}
Taking into account the recurrence relation \reff{13-1.1},
we define the "position operators "
$Z$ and  $\overline{Z}$ by the relations
\begin{subequations}   \begin{eqnarray}
Z\,U_{k,\,l}(z,\,\z)=U_{k+1,\,l}(z,\,\z)+U_{k,\,l-1}(z,\,\z)+U_{k-1,\,l+1}(z,\,\z),
\label{13-2.1}\\
\overline{Z}\,U_{k,\,l}(z,\,\z)=
U_{k,\,l+1}(z,\,\z)+U_{k-1,\,l}(z,\,\z)+U_{k+1,\,l-1}(z,\,\z). \label{13-2.2}
\end{eqnarray} \end{subequations}
These relations define the operators $Z$ and $\overline{Z}$ on the dense in $\H$
linear span of all ChK-polynomials $U_{k,\,l}(z,\,\z)$.
After closing of these operators they become  bounded operators (which we denote by the
same symbols) on the whole space $\H$ and satisfy the  relations
$Z^*=\overline{Z}$, $\overline{Z}^*=Z$.

Following \cite{N16}, we define the "momentum operators" as operators that
conjugate relatively  basis in $\H$ to the above position operators.
For this purpose we consider the integral operator $\mathbb{K}$ with the Poisson kernel
\begin{equation}\label{13-2.4}
K_s(z,\overline{z};\zeta,\overline{\zeta})=
\sum_{k,\,l=0}^{\infty}(-1)^{k+l}
\overline{U_{k,\,l}(z,\,\z)}U_{k,\,l}(\zeta,\,\overline{\zeta}).
\end{equation}
The operator $\mathbb{K}$ acts from the Hilbert space
$\H=\text{L}^2(S,\mu(\text{d}x,\text{d}y)$\, ($z=x+iy$) into the Hilbert space
$\H_1=\text{L}^2(S,\mu(\text{d}\zeta,\text{d}\overline{\zeta})$\, ($\zeta=\xi+i\eta$).
In the space $\H_1$ we define "position operators" $\lz$ и $\lzz$, using the
formulas similar
to \reff{13-2.1} и \reff{13-2.2}.
For the inverse operator $\mathbb{K}^{-1}:\H_1\rightarrow\H$ is fulfilled the relation $\mathbb{K}^{-1}=\mathbb{K}^{*}$, where $\mathbb{K}^{*}$ is the integral operator
adjoint to $\mathbb{K}$. Hence $\mathbb{K}$ is an unitary operator.

Using the operator $\mathbb{K}$, we define "momentum operators" $P_Z=Z^\dag$
and $P_{\overline{Z}}=\overline{Z}^\dag$, conjugate to $Z$ and $\overline{Z}$
with respect to selected basis of the space $\H$ :
\begin{equation}\label{13-2.3}
   Z^\dag=\mathbb{K}^{-1}\lz\mathbb{K},\qquad
   \overline{Z}^\dag=\mathbb{K}^{-1}\lzz\mathbb{K}.
\end{equation}
From \reff{13-2.3} it follows that
operators $P_Z=Z^\dag$ and $P_{\overline{Z}}=\overline{Z}^\dag$ are  bounded
in $\H$ and adjoint to each other
\begin{equation*}
   (Z^\dag)^*=\overline{Z}^\dag, \qquad (\overline{Z}^\dag)^*=Z^\dag.
\end{equation*}
The action of the operators $Z^\dag$ and $\overline{Z}^\dag$ on the basic elements of the space
$\H$ is given by formulas
\begin{subequations}   \begin{eqnarray}
Z^\dag\,U_{k,\,l}(z,\,\z)=-U_{k+1,\,l}(z,\,\z)-U_{k,\,l-1}(z,\,\z)+U_{k-1,\,l+1}(z,\,\z),
\label{13-2.5}\\
\overline{Z}^\dag\,U_{k,\,l}(z,\,\z)=
-U_{k,\,l+1}(z,\,\z)-U_{k-1,\,l}(z,\,\z)+U_{k+1,\,l-1}(z,\,\z). \label{13-2.6}
\end{eqnarray} \end{subequations}
Now we define the quadratic Hamiltonian of sectorial oscillator as
\begin{equation*}
    H_s=H_s^{(1)}+H_s^{(2)},
\end{equation*}
where
\begin{equation*}
H_s^{(1)}=\frac{1}{40}
\left(Z\overline{Z}+Z\overline{Z}^{\dag}+Z^{\dag}\overline{Z}+Z^{\dag}\overline{Z}^{\dag}\right);
\quad H_s^{(2)}=\frac{1}{40}
\left(\overline{Z}Z+\overline{Z}Z^{\dag}+\overline{Z}^{\dag}Z+\overline{Z}^{\dag}Z^{\dag}\right).
\end{equation*}
Hamiltonian $H_s$ is a self-adjoint operator in the space $\H$.
ChK-polynomials  $U_{k,\,l}(z,\,\z)$
are the eigenfunctions of $H_s$ with eigenvalues
\begin{equation*}
\lambda_{0,0}=0;\, \lambda_{N,0}=\lambda_{0,N}=\frac{1}{10}, \,(N\geq1);\,
\lambda_{k,\,l}=\frac{1}{5},\, (k,l\geq1).
\end{equation*}
We consider $\H$ as a Fock space
and define ladder operators
$\aspm$ by the relations
\begin{equation*}
\asp=\frac{1}{\sqrt{40}}\left(Z+Z^{\dag}\right),\quad
\asm=\frac{1}{\sqrt{40}}\left(\overline{Z}+\overline{Z}^{\dag}\right).
\end{equation*}
These operators acting in "$N$-particle" subspaces  $\H^{(N)}$ ($N=k+l$) as follows
\begin{equation}\label{13-2.10}
\aspm U_{k,\,l}(z,\,\z)=\frac{1}{\sqrt{10}}U_{k\mp 1,\,l\pm 1}(z,\,\z),\qquad
\asp U_{0,N}(z,\,\z)=0,\quad \asm U_{N,0}(z,\,\z)=0.
\end{equation}
Using \reff{13-1.2}, we have shown in \cite{N47} that
the ladder operators can be represented by the differential operators:
\begin{equation*}
\aspm=\frac{1}{\sqrt{10}}\bigoplus_{N=1}^{\infty}\sum_{m=0}^{N}
U_{m\mp 1,\,N-(m\mp 1)}\frac{\mathbf{D}_{m,\,N-m}}{m!(N-m)!},
\end{equation*}
where
\begin{equation*}
\mathbf{D}_{m,\,N-m}=\frac{\partial^N}{\partial z^m \partial \z^{N-m}}.
\end{equation*}
The ladder operators are adjoint to each other
$\left(\aspm)\right)^*=\asmp$,
and Hamiltonian in terms of these operators has the form
\begin{equation*}
H_s=H_s^{(1)}+H_s^{(2)},\qquad
H_s^{(1)}=a_{sect}^{+}a_{sect}^{-},\quad H_s^{(2)}=a_{sect}^{-}a_{sect}^{+}.
\end{equation*}
Self-adjoint "number" operators $\mathbb{N}_1$,\, $\mathbb{N}_2$ are defined
by their action on the basic elements of the space $\H$:
\begin{equation*}
\mathbb{N}_1U_{k,\,l}=kU_{k,\,l},\quad
\mathbb{N}_2U_{k,\,l}=lU_{k,\,l}.
\end{equation*}
We also need two auxiliary operators
\begin{equation}\label{13-2.15}
\mathbb{P}_1=\bigoplus_{N=0}^{\infty}P_{N,\,0},\qquad \mathbb{P}_2=
\bigoplus_{N=0}^{\infty}P_{0,\,N},
\end{equation}
where
$$
P_{N,\,0}U_{k,\,l}(z,\,\z)=\delta_{k,\,N}\delta_{l,\,0}U_{k,\,l}(z,\,\z),\quad
P_{0,\,N}U_{k,\,l}(z,\,\z)=\delta_{k,\,0}\delta_{l,\,N}U_{k,\,l}(z,\,\z).
$$

In \cite{N46} the algebra $\mathfrak{A}_s$ of generalized sectorial oscillator
was defined as the closure of an associative algebra generated by operators
\begin{equation}\label{13-2.16}
\mathbb{I},\, a_{sect}^{\pm},\, \mathbb{N}_1,\, \mathbb{N}_2,\, \mathbb{P}_1,\, \mathbb{P}_2,
\end{equation}
satisfying the commutation relations
\begin{gather}
[a_{sect}^{+},a_{sect}^{-}]=\frac{1}{10}\left(\mathbb{P}_1-\mathbb{P}_2\right);\quad
[\mathbb{N}_1,a_{sect}^{\pm}]=\mp a_{sect}^{\pm};\quad
[\mathbb{N}_2,a_{sect}^{\pm}]=\pm a_{sect}^{\pm};\nn\\
[\mathbb{P}_1,\mathbb{P}_2]=0,\quad [\mathbb{N}_1,\mathbb{N}_2]=0,\quad
[\mathbb{N}_i,\mathbb{P}_j]=0,\,(i,j=1,2);\label{13-2.17}\\
\mathbb{P}_1a_{sect}^{+}=0,\quad a_{sect}^{-}\mathbb{P}_1=0,\quad
a_{sect}^{+}\mathbb{P}_2=0,\quad \mathbb{P}_2a_{sect}^{-}=0.\nn
\end{gather}

\subsubsection{Generalized radial oscillator }

As in the case of sectorial oscillator, we start  by defining  the  "position" operator
\begin{equation}\label{13-2.18}
X:=\!-5H_s\!-\!\frac{1}{4}\left(Z\Z^{\,*}+Z^*\Z+\Z Z^*+\Z^{\,*}Z\right),\quad \text{Dom}[X]\!=\!\H.
\end{equation}
By analogy with \reff{13-2.3}, using the Poisson kernel
\begin{equation}\label{13-2.19}
K_r(z,\overline{z};\zeta,\overline{\zeta})=
\sum_{k,\,l=0}^{\infty}e^{i\frac{\pi}{4}(k+l)}
\overline{U_{k,\,l}(z,\,\z)}U_{k,\,l}(\zeta,\,\overline{\zeta}),
\end{equation}
we define the momentum operator $P_X=X^\dag$ conjugate relative basis in the space $\H$
to the coordinate operator. The quadratic Hamiltonian of radial oscillator
\begin{equation}\label{13-2.20}
H_r=\frac{1}{4}\left(X^2+(X^*)^2\right),\quad \text{Dom}[H_r]=\H,
\end{equation}
is a bounded selfadjoint operator in $\H$.
ChK-polynomials  $U_{k,\,l}(z,\,\z)$
are eigenfunctions of the operator  $H_r$ with eigenvalues
\begin{equation}\label{13-2.21}
\nu_{0\,0}=\frac{4}{5},\quad  \nu_{N\,0}=\nu_{0\,N}=\frac{1}{2}\,\, (\text{for}\, N\geq1),
\quad \nu_{m\,n}=\frac{1}{5}\,\, (\text{for}\, n,m\geq1).
\end{equation}
To determine the ladder operators we introduce an auxiliary operator
$I(\alpha,\beta)$\quad ( $\alpha,\beta\in \mathbb{C}$) which acts
on the basic elements $U_{k,\,l}(z,\,\z)$ as follows
\begin{equation}\label{13-2.22}
I_B(\alpha;\beta)U_{k,\,l}(z,\,\z)=
\left\{\alpha\left(\delta_{k,N}\delta_{l,0}+\delta_{k,0}\delta_{l,N}\right)
\left(1-\delta_{k,0}\delta_{l,0}\right)+\beta\delta_{k,0}\delta_{l,0}\right\}U_{k,\,l}(z,\,\z).
\end{equation}
Using this operator, we define the ladder operators of radial oscillator by the relations
\begin{equation}\label{13-2.23}
\arpm=\frac{1}{\sqrt{10}}\left(X+iX^\dag\right)-\frac{2}{\sqrt{5}}\,
I_B\left(\frac{1}{4}\,e^{\pm i\frac{\pi}{4}},\frac{1}{2}\,e^{\pm i\frac{\pi}{4}}\right).
\end{equation}
From \reff{13-2.18},\reff{13-2.22},\reff{13-2.23} it follows that the action of $\arpm$ looks as
\begin{equation}\label{13-2.24}
\arpm U_{k,\,l}(z,\,\z)=\sqrt{\frac{2}{5}}\,U_{k\pm 1,\,l\pm 1}(z,\,\z).
\end{equation}
In the work \cite{N47} we have found  the following differential operator representation of
the operators $\arpm$:
\begin{equation*}
a_{rad}^{\pm}\,\rule[-10pt]{.2pt}{20pt}_{\,\H^{(N)}}=\sqrt{\frac{2}{5}}\,
\sum_{m=0}^{N}U_{m\pm 1,\,N-m\pm 1}\frac{\mathbf{D}_{m,\,N-m}}{m!(N-m)!}.
\end{equation*}
The (position) operator is given by the relation
\begin{equation*}
a_{rad}^{-}+a_{rad}^{+}=\sqrt{\frac{2}{5}}\left\{\rule[-5pt]{0pt}{22pt}
(z\z -3)I-\bigoplus_{N=1}^{\infty}\left[\!\sum_{m=0}^N\frac{\alpha_1}{m!(N-m)!}
-\sum_{m=0}^{N-1}\frac{\alpha_2}{m!(N-m)!}\right]\right\},
\end{equation*}
where
\begin{gather*}
\alpha_1=U_{m-1,\,N-m+1}\mathbf{D}_{m,N-m}\overline{Z}+
U_{m+1,\,N-m-1}\mathbf{D}_{m,N-m}Z; \\
\alpha_2=U_{m,\,N-m}\mathbf{D}_{m,N-m}+
U_{N-m,\,m}\mathbf{D}_{N-m,m},
\end{gather*}
and $Z$ , $\overline{Z}$ are
the operators of multiplication by  $z$ and $\z$, respectively.
From \reff{13-2.18}-\reff{13-2.20}, \reff{13-2.22}-\reff{13-2.24} we obtain the
following expression for the Hamiltonian $H_r$ in terms of the ladder operators
\begin{equation}\label{13-2.28}
H_r=\arp\arm +\arm\arp +I_B\left(\ts\frac{1}{8};\ts\frac{1}{2}\right).
\end{equation}
Note that, unlike the Hamiltonian $H_s$  sectorial oscillator, the Hamiltonian $H_r$ radial oscillator
contains an extra term $I_B\left(\ts\frac{1}{8};\ts\frac{1}{2}\right)$ in
addition to standard members.

In the work  \cite{N46} the algebra $\mathfrak{A}_r$ of generalized radial
oscillator was defined as a closure of the associative algebra generated by operators
\begin{equation}\label{13-2.29}
a^{\pm}_{rad},\quad \mathbb{N}_1,\quad \mathbb{N}_2,\quad I_B(\ts\half,\ts\half),
\quad \mathbb{I},
\end{equation}
satisfying the following commutation relations
\begin{gather}
[a^{-}_{rad},a^{+}_{rad}]=\frac{4}{5}I_B(\ts\half,\ts\half);
\quad a^{-}_{rad}I_B(\ts\half,\ts\half)=0;\quad I_B(\ts\half,\ts\half)a^{+}_{rad}=0;\nn\\
[\mathbb{N}_1,a^{\pm}_{rad}]=\pm a^{\pm}_{rad};\quad
[\mathbb{N}_2,a^{\pm}_{rad}]=\pm a^{\pm}_{rad};\quad
[\mathbb{N}_1,I_B(\ts\half,\ts\half)]=0,\quad
[\mathbb{N}_2,I_B(\ts\half,\ts\half)]=0. \label{13-2.30}
\end{gather}

\subsubsection{Generalized boundary oscillator }
Recall that  $\Hco$ and $\Hoc$ denote the closures in $\H$ of linear spans of the sets
$\bask$ and $\basl$.
To define the boundary  oscillator we use \cite{N46} the decomposition of the
Hilbert spaces $\Hco$ and $\Hoc$
\begin{equation}\label{13-2.31}
\Hco=\bigoplus_{N=0}^{\infty}\left(\H^{(N)}\cap\Hco\right)\quad
\Hoc=\bigoplus_{N=0}^{\infty}\left(\H^{(N)}\cap\Hoc\right)
\end{equation}
and operators
$P^{m,\,n}_{s,\,t}\, \suchthat\, P^{m,\,n}_{s,\,t}U_{k,\,l}=
\delta_{k,\,m}\delta_{l,\,n}U_{s,\,t}$.
The ladder operators are defined  by decompositions (see Fig.2)
\begin{equation}\label{13-2.32}
\acopm=\bigoplus_{m=0}^{\infty}P_{m\pm 1,\,0}^{m,\,0},\quad
\aocpm=\bigoplus_{m=0}^{\infty}P_{0,\,m\pm 1}^{0,\,m},
\end{equation}
and have \cite{N47} the following differential operator representation:
\begin{equation*}
\acopm\rule[-10pt]{.2pt}{20pt}_{\,\,\H^{(N)}}
=U_{N\pm 1,\,0}\frac{\mathbf{D}_{N,\,0}}{N!};\quad
\aocpm\rule[-10pt]{.2pt}{20pt}_{\,\,\H^{(N)}}
=U_{0,\,N\pm 1}\frac{\mathbf{D}_{0,\,N}}{N!}.
\end{equation*}
For position operators
\begin{align*}
\acop+\acom&=\left(Z+\overline{Z}-a_{sect}s^+-a_{sect}^+Z+
(a_{sect}^+)^2\right)\mathbb{P}_1,\nn \\
\aocp+\aocm&=\left(Z+\overline{Z}-a_{sect}^--a_{sect}^-Z+
(a_{sect}^-)^2\right)\mathbb{P}_2,
\end{align*}
this gives \cite{N47} the following representation
by differential operators:
\begin{multline*}
\acop+\acom=Z+\overline{Z}-\frac{1}{\sqrt{10}} \times
\left\{ \bigoplus_{N=1}^{\infty} \left[\left(
\sum_{m=0}^N \frac{U_{m-1,\,N-m+1}\,\mathbf{D}_{m,\,N-m}}{m!(N-m)!}
\right)(\mathbb{I}+Z)\right.\right.
\\
-\left.\left.
\left(\sum_{m=0}^{N-2}
\frac{U_{m,\,N-m}\,\mathbf{D}_{m+2,\,N-m-2}}{(m+2)!(N-m-2)!}\right)
\right]\right\}\mathbb{P}_1;
\end{multline*}
\begin{multline*}
\aocp+\aocm=Z+\overline{Z}-\frac{1}{\sqrt{10}} \times
\left\{\bigoplus_{N=1}^{\infty}\left[\left(
\sum_{m=0}^N \frac{U_{m+1,\,N-m-1}\,\mathbf{D}_{m,\,N-m}}{m!(N-m)!}
\right)(\mathbb{I}+\overline{Z})\right.\right. \\
-\left.\left.\left(\sum_{m=0}^{N-2}
\frac{U_{m+2,\,N-m-2}\,\mathbf{D}_{m,\,N-m}}{m!(N-m)!}\right)
\right]\right\}\mathbb{P}_2.
\end{multline*}
The quadratic Hamiltonian of the boundary oscillator has the form
\begin{equation*}
H_0=\frac15\left(H_{\bullet,\,0}+H_{0,\,\bullet}-P^{0,\,0}_{0,\,0}\right),
\end{equation*}
where
\begin{align*}
H_{\bullet,\,0}=
\left(a^{+}_{\bullet,\,0}a^{-}_{\bullet,\,0}+a^{-}_{\bullet,\,0}a^{+}_{\bullet,\,0}\right),\\
H_{0,\,\bullet}=
\left(a^{+}_{0,\,\bullet}a^{-}_{0,\,\bullet}+a^{-}_{0,\,\bullet}a^{+}_{0,\,\bullet}\right).
\end{align*}
The Hamiltonian $H_0$ is a bounded selfadjoint operator in $\H$.
ChK-polynomials  $U_{k,\,l}(z,\,\z)$
are eigenfunctions for $H_0$  with eigenvalues
\begin{equation*}
\mu_{0,\,0}=\frac{1}{5};\quad\mu_{0,\,N}=\mu_{N,\,0}=\frac{2}{5}, \,(\text{for}\, N\geq1),
\quad \mu_{k,\,l}=0, \,(\text{for}\, k,l\geq1).
\end{equation*}
Then the algebra  $\mathfrak{A}_{0}$ of generalized boundary oscillator
was defined in \cite{N46} as a closure of the associative algebra with generators
\begin{equation}\label{13-2.36}
\mathbb{I},\quad \mathbb{N}_1,\quad \mathbb{N}_2,\quad \Pnkm{s,\,0}{m}{0},
\quad \Pnkm{0,\,s}{0}{m},\quad \Pnkm{k,\,0}{0}{m},\quad \Pnkm{0,\,k}{m}{0},
\end{equation}
satisfying the commutation relations
\begin{equation}\label{13-2.37}
\left\{
\begin{aligned}
\left[\mathbb{N}_1,\mathbb{N}_2\right]=0,
\quad [\mathbb{N}_1,\Pnkm{k,\,0}{m}{0}]=(m-k)\Pnkm{k,\,0}{m}{0},\quad
[\mathbb{N}_1,\Pnkm{0,\,k}{0}{m}]=0,& \\
[\mathbb{N}_1,\Pnkm{k,\,0}{0}{n}]=-k\Pnkm{k,\,0}{0}{n},\quad
[\mathbb{N}_1,\Pnkm{0,\,l}{m}{0}]=m\Pnkm{0,\,l}{m}{0};\qquad\qquad&
\end{aligned}
\right.
\end{equation}

\begin{equation}\label{13-2.38}
\left\{
\begin{aligned}
\left[\mathbb{N}_2,\Pnkm{k,\,0}{m}{0}\right]=0,
\quad \left[\mathbb{N}_2,\Pnkm{0,\,k}{0}{m}\right]=(m-k)\Pnkm{0,\,k}{0}{m},& \\
\left[\mathbb{N}_2,\Pnkm{k,\,0}{0}{n}\right]=n\Pnkm{k,\,0}{0}{n},\quad
\left[\mathbb{N}_2,\Pnkm{0,\,l}{m}{0}\right]=-l\Pnkm{0,\,l}{m}{0};&
\end{aligned}
\right.
\end{equation}

\begin{equation}\label{13-2.39}
\left\{
\begin{aligned}
\left[\Pnkm{k,\,0}{m}{0},\Pnkm{l,\,0}{n}{0}\right]&=
\delta_{k,\,n}\Pnkm{l,\,0}{m}{0}-\delta_{m,\,l}\Pnkm{k,\,0}{n}{0}, \\
\left[\Pnkm{k,\,0}{m}{0},\Pnkm{0,\,l}{0}{n}\right]&=
\delta_{k,\,0}\delta_{n,\,0}\Pnkm{0,\,l}{m}{0}-
\delta_{m,\,0}\delta_{l,\,0}\Pnkm{k,\,0}{0}{n};
\end{aligned}
\right.
\end{equation}

\begin{equation}\label{13-2.40}
\left\{
\begin{aligned}
\left[\Pnkm{k,\,0}{m}{0},\Pnkm{l,\,0}{0}{n}\right]&=
\delta_{k,\,0}\delta_{n,\,0}\Pnkm{l,\,0}{m}{0}-
\delta_{l,\,m}\Pnkm{k,\,0}{0}{n}, \\
\left[\Pnkm{k,\,0}{m}{0},\Pnkm{0,\,l}{n}{0}\right]&=
\delta_{k,\,n}\Pnkm{0,\,l}{m}{0}-
\delta_{m,\,0}\delta_{l,\,0}\Pnkm{k,\,0}{n}{0};
\end{aligned}
\right.
\end{equation}

\begin{equation}\label{13-2.41}
\left\{
\begin{aligned}
\left[\Pnkm{0,\,k}{0}{m},\Pnkm{0,\,l}{0}{n}\right]&=
\delta_{k,\,n}\Pnkm{0,\,l}{0}{m}-\delta_{m,\,l}\Pnkm{0,\,l}{0}{n}, \\
\left[\Pnkm{0,\,k}{0}{m},\Pnkm{l,\,0}{0}{n}\right]&=
\delta_{k,\,n}\Pnkm{l,\,0}{0}{m}-
\delta_{m,\,0}\delta_{l,\,0}\Pnkm{0,\,k}{0}{n}, \\
\left[\Pnkm{0,\,k}{0}{m},\Pnkm{0,\,l}{n}{0}\right]&=
\delta_{k,\,0}\delta_{n,\,0}\Pnkm{0,\,l}{0}{m}-
\delta_{l,\,m}\Pnkm{0,\,k}{n}{0};
\end{aligned}
\right.
\end{equation}

\begin{equation}\label{13-2.42}
\left\{
\begin{aligned}
\left[\Pnkm{k,\,0}{0}{n},\Pnkm{0,\,l}{m}{0}\right]&=
\delta_{k,\,m}\Pnkm{0,\,l}{0}{n}-
\delta_{l,\,n}\Pnkm{k,\,0}{m}{0}, \\
\left[\Pnkm{k,\,0}{0}{n},\Pnkm{l,\,0}{0}{m}\right]&=
\delta_{k,\,0}\delta_{m,\,0}\Pnkm{l,\,0}{0}{n}-
\delta_{l,\,0}\delta_{n,\,0}\Pnkm{k,\,0}{0}{m}, \\
\left[\Pnkm{0,\,k}{n}{0},\Pnkm{0,\,l}{m}{0}\right]&=
\delta_{k,\,0}\delta_{m,\,0}\Pnkm{0,\,l}{n}{0}-
\delta_{l,\,0}\delta_{k,\,0}\Pnkm{0,\,k}{m}{0}.
\end{aligned}
\right.
\end{equation}

\section{Algebra of generalized sectorial oscillator}

In this section, we define the infinite-dimensional Lie algebra $\mathfrak{A}_s$
as a closure of the sectorial oscillator algebra which is the (minimal)
expansion of an associative algebra determined \reff{13-2.16} and \reff{13-2.17}.
To do this we need some auxiliary operators entered below.
Recall that
\begin{equation}\label{13-3.1}
P^{m,\,n}_{s,\,t}U_{k,\,l}(z,\,\z)=\delta_{k,\,m} \delta_{l,\,n}U_{s,\,t}(z,\,\z),
\end{equation}
if $m,n,s,t,k,l\geq0$, and
\begin{equation*}
P^{m,\,n}_{s,\,t}U_{k,\,l}(z,\,\z)=0,
\end{equation*}
if at least one of the indices is negative.

Using \reff{13-3.1}, \reff{13-2.10} and decomposition \reff{13-1.2}, we have
\begin{equation}\label{13-3.2}
\aspm=\frac{1}{\sqrt{10}}\bigoplus_{m,\,n=0}^{\infty}P^{m,\,n}_{m\mp 1,\,n\pm 1}.
\end{equation}
We introduce the operators
\begin{equation}\label{13-3.3}
P^{(n)}_{\bullet,\,k}=\bigoplus_{m=k}^{\infty}P^{m-k,\,k}_{m-n,\,n},
\qquad P^{(q)}_{p,\,\bullet}=\bigoplus_{l=p}^{\infty}P^{p,\,l-p}_{q,\,l-q},
\qquad k,n,q,p\geq0.
\end{equation}
The relations \reff{13-2.15} and \reff{13-3.3} imply that
\begin{equation*}
P_1=\bigoplus_{N=0}^{\infty}P^{N,\,0}_{N,\,0}=P^{(0)}_{\bullet,\,0},
\qquad P_2=\bigoplus_{N=0}^{\infty}P^{0,\,N}_{0,\,N}=P^{(0)}_{0,\,\bullet}.
\end{equation*}
The action of all operators defined above shown in the following diagram.
\bigskip
\begin{figure}[h]
\centering
\includegraphics[width=12.0cm]{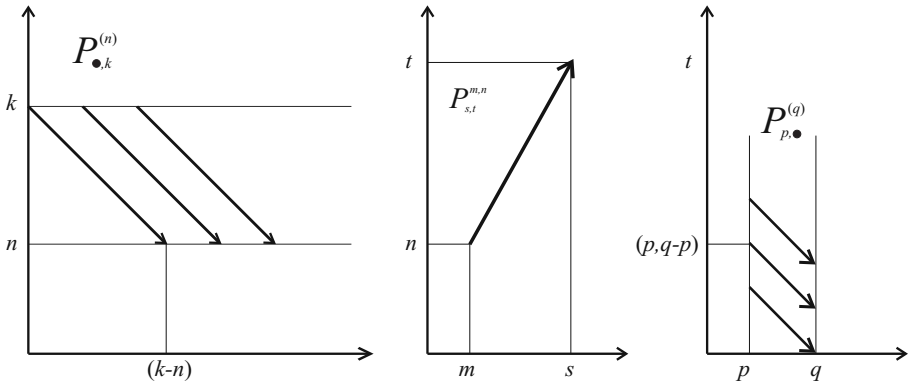}
\caption{The action of operators \reff{13-3.1}, \reff{13-3.3}}
\end{figure}

We will start building algebra $\mathfrak{A}_s$ of the sectorial oscillator
with the definition of algebras $\mathfrak{A}^{(N)}_s$,\, for $N\geq0$,
as associative algebras with generators
\begin{equation}\label{13-3.4}
\mathbb{I},\, \mathbb{N}_1,\, \mathbb{N}_2,\, P^{k,\,l}_{k\mp 1,\,l\pm 1},
\, P^{k,\,l}_{k,\,l},\quad
(k+l=N,\, k,l\geq0),
\end{equation}
satisfying the commutation relations
\begin{gather}
[\mathbb{N}_1,\,\mathbb{N}_2]=0,
\quad [\mathbb{N}_1,\,P^{k,\,l}_{k,\,l}]=[\mathbb{N}_2,\,P^{k,\,l}_{k,\,l}]=0,\quad
[P^{k,\,l}_{k\,l},\,P^{m,\,n}_{m\,n}]=0,\nn\\
[\mathbb{N}_1,\,P^{k,\,l}_{k\mp 1,\,l\pm 1}]=\mp P^{k,\,l}_{k\mp 1,\,l\pm 1},\qquad
[\mathbb{N}_2,\,P^{k,\,l}_{k\mp 1,\,l\pm 1}]=
\pm P^{k,\,l}_{k\mp 1,\,l\pm 1},\label{13-3.5}\\
[P^{m,\,n}_{m\mp 1,\,n\pm 1},\,P^{k,\,l}_{k,\,l}]=
\delta_{m,\,k}\delta_{n,\,l}P^{k,\,l}_{m\mp 1,\,n\pm 1}
-\delta_{k,\,m\mp 1}\delta_{l,\,n\pm 1}P^{m,\,n}_{k,\,l},\nn\\
[P^{m,\,n}_{m-1,\,n+1},\,P^{k,\,l}_{k+1,\,l-1}]=\delta_{m,\,k+1}\delta_{n,\,l-1}
\left(P^{k,\,l}_{k,\,l}-P^{k+1,\,l-1}_{k+1,\,l-1}\right).
\nn
\end{gather}
It is easy to check the validity of the Jacobi identities.
Therefore, the algebras $\mathfrak{A}^{(N)}_s,$ $N\geq0$, generated by operators
\reff{13-3.4} satisfying  the relations \reff{13-3.5}, are  Lie algebras with
dimension  $3(N+2)$. Using the decomposition \reff{13-1.2}, we determine the
infinite-dimensional Lie algebra $\widetilde{\mathfrak{A}}_s$
\begin{equation*}
\AGts=\bigoplus_{N=0}^{\infty}\AG^{(N)}_s,
\end{equation*}
We define the Lie algebra $\widehat{\AG}_s$ as algebra,
obtained from $\widetilde{\AG}_s$ by addition to generators \reff{13-3.4}
of all formal series composed of these generators, including operators \reff{13-3.2} and \reff{13-3.3}.
The commutation relations of the algebra $\widehat{\AG}_s$ are induced  by the relations \reff{13-3.5}.

Let  $I_s$ be an ideal of the algebra $\widehat{\AG}_s$ generated by commutation relations:
\begin{equation}\label{13-3.7}
[\asm,\asp]=\frac{1}{10} \left(\Poco - \Pooc\right);
\end{equation}
\begin{equation}\label{13-3.8}
[\mathbb{N}_1,\aspm]=\mp\aspm,\quad [\mathbb{N}_2,\aspm]=\pm\aspm,
\quad [\mathbb{N}_1,\mathbb{N}_2]=0;
\end{equation}
\begin{equation}\label{13-3.9}
[\aspm,\Pnck{n}{k}]=\frac{1}{\sqrt{10}}\left(\Pnck{n\pm 1}{k}-\Pnck{n}{k\pm 1}\right);
\end{equation}
\begin{equation}\label{13-3.10}
[\aspm,\Pnkc{n}{k}]=\frac{1}{\sqrt{10}}\left(\Pnkc{n\mp 1}{k}-\Pnkc{n}{k\pm 1}\right);
\end{equation}
\begin{equation}\label{13-3.11}
[\asmp,\Pnkm{k,\,l}{k-t}{l+t}]=
\frac{1}{\sqrt{10}}\left(\Pnkm{k,\,l}{k-t\pm 1}{l+t\mp 1}-
\Pnkm{k\mp 1,\,l\pm 1}{k-t}{l+t}\right);
\end{equation}
\begin{equation}\label{3a.12}
[\asmp,\Pnkm{k,\,l}{k+t}{l-t}]=
\frac{1}{\sqrt{10}}\left(\Pnkm{k,\,l}{k+t\pm 1}{l-t\mp 1}-
\Pnkm{k\mp 1,\,l\pm 1}{k+t}{l-t}\right);
\end{equation}
\begin{equation}\label{13-3.13}
[\Pnck{n}{k},\Pnkc{m}{l}]=\left(\Pnkm{l,\,m+k-l}{m+k-n}{n}-\Pnkm{n+l-k,\,k}{m}{n+l-m}\right);
\end{equation}
\begin{equation}\label{13-3.14}
[\Pnck{n}{k},\Pnck{m}{l}]=\Pnck{n}{l}-\Pnck{m}{k};
\end{equation}
\begin{equation}\label{13-3.15}
[\Pnkc{n}{k},\Pnkc{m}{l}]=\Pnkc{n}{l}-\Pnkc{m}{k};
\end{equation}
\begin{equation}\label{13-3.16}
[\mathbb{N}_1,\Pnck{n}{k}]=(k-n)\Pnck{n}{k},\qquad
[\mathbb{N}_2,\Pnck{n}{k}]=(n-k)\Pnck{n}{k};
\end{equation}
\begin{equation}\label{13-3.17}
[\mathbb{N}_1,\Pnkc{n}{k}]=(n-k)\Pnkc{n}{k},
\qquad [\mathbb{N}_2,\Pnkc{n}{k}]=(k-n)\Pnkc{n}{k};
\end{equation}
\begin{equation}\label{13-3.18}
[\Pnkm{k,\,l}{k-t}{l+t},\Pnck{n}{m}]=
\Pnkm{k+n-m,\,m}{k-t}{n+t}-\Pnkm{k,\,l}{k+l-n}{n};
\end{equation}
\begin{equation}\label{13-3.19}
[\Pnkm{k,\,l}{k+t}{l-t},\Pnck{n}{m}]=
\Pnkm{k+l-m,\,m}{k+t}{l-t}-\Pnkm{k,\,l}{k+l-n}{n};
\end{equation}
\begin{equation}\label{13-3.20}
[\Pnkm{k,\,l}{k+t}{l-t},\Pnkc{n}{m}]=
\Pnkm{m,\,k+l-m}{k+t}{l-t}-\Pnkm{k,\,l}{n}{k+l-n};
\end{equation}
\begin{equation}\label{13-3.21}
[\Pnkm{k,\,l}{k-t}{l+t},\Pnkc{n}{m}]=
\Pnkm{m,\,k+l-m}{k-t}{l+t}-\Pnkm{k,\,l}{n}{k+l-n};
\end{equation}
\begin{equation}\label{13-3.22}
[\Pnkm{k,\,l}{k\mp t}{l\pm t},\Pnkm{m,\,n}{m\mp s}{n\pm s}]=
\delta_{k,\,m\mp s}\delta_{l,\,n\pm s}\Pnkm{m,\,n}{k\mp t}{l\pm t}-
\delta_{m,\,k\mp t}\delta_{n,\,l\pm t}\Pnkm{k,\,l}{m\mp s}{n\pm s};
\end{equation}
\begin{equation}\label{13-3.23}
[\Pnkm{k,\,l}{k-t}{l+t},\Pnkm{m,\,n}{m+s}{n-s}]=
\delta_{k,\,m+s}\delta_{l,\,n-s}\Pnkm{m,\,n}{k-t}{l+t}-
\delta_{m,\,k-t}\delta_{n,\,l+t}\Pnkm{k,\,l}{m+s}{n-s};
\end{equation}
\begin{equation}\label{13-3.24}
[\mathbb{N}_1,\Pnkm{k,\,l}{k\mp t}{l\pm t}]=\mp t\Pnkm{k,\,l}{k\mp t}{l\pm t},\qquad
[\mathbb{N}_2,\Pnkm{k,\,l}{k\mp t}{l\pm t}]=\pm t\Pnkm{k,\,l}{k\mp t}{l\pm t}.
\end{equation}

Finally, we define the algebra $\AG_s$ of sectorial oscillator as quotient algebra
\begin{equation}\label{13-3.25}
\AG_s=\widehat{\AG}_s/I_s.
\end{equation}
In other words, algebra $\AG_s$ is Lie algebra with generators
\reff{13-3.2}, \reff{13-3.3}, \reff{13-3.4}
satisfy the commutation relations \reff{13-3.7}-\reff{13-3.24}.
Clearly, that $\AG_s$ is a closure of the associative algebra  generated
by operators \reff{13-2.16} and relations \reff{13-2.17}.

\section{Algebra of generalized radial oscillator}

In this section, we define the infinite-dimensional Lie algebra  $\AG_r$
of the radial oscillator just as for sectorial oscillator. Namely, we
describe the (minimal) extension of an associative
algebra determined by generators \reff{13-2.29} with commutation relations \reff{13-2.30}.

Using relations \reff{13-3.1} and \reff{13-2.24}, we obtain
\begin{equation}\label{13-4.1}
\arpm=\sqrt{\frac{2}{5}}\bigoplus_{k,\,l\geq0}\Pnkm{k,\,l}{k\pm 1}{l\pm 1}.
\end{equation}
From  \reff{13-3.1}, \reff{13-3.3} and \reff{13-2.22} should be  the following equality
\begin{equation}\label{13-4.2}
I_B(\frac{1}{2},\frac{1}{2})=\frac{1}{2}\left(\Pnck{0}{0}+\Pnkc{0}{0}-\Pnkm{0,\,0}{0}{0}\right).
\end{equation}
Let's define algebras $\AGte$ and $\AGto$ as associative algebras with generators
\begin{equation}\label{13-4.3}
\mathbb{I},\, \mathbb{N}_1,\, \mathbb{N}_2,\, \Pnkm{k,\,l}{k\pm t}{l\pm t}, \quad k,l,t\geq 0,
\end{equation}
(where $k=2n,\,l=2m$ и $k=2n+1,\,l=2m+1$, respectively)
satisfying commutation relations
\begin{equation}\label{13-4.4}
[\Pnkm{k,\,l}{k\pm t}{l\pm t},\,\Pnkm{m,\,n}{m\pm s}{n\pm s}]=
\delta_{k,\,m\pm s}\delta_{l,\,n\pm s}\Pnkm{m,\,n}{k\pm t}{l\pm t}-
\delta_{m,\,k\pm t}\delta_{n,\,l\pm t}\Pnkm{k,\,l}{m\pm s}{n\pm s};
\end{equation}
\begin{equation}\label{13-4.5}
[\Pnkm{k,\,l}{k+ t}{l+ t},\,\Pnkm{m,\,n}{m-s}{n-s}]=
\delta_{k,\,m-s}\delta_{l,\,n-s}\Pnkm{m,\,n}{k+t}{l+t}-
\delta_{m,\,k+t}\delta_{n,\,l+t}\Pnkm{k,\,l}{m-s}{n-s};
\end{equation}
\begin{equation}\label{13-4.6}
[\mathbb{N}_1,\Pnkm{k,\,l}{k\pm t}{l\pm t}]=\pm t \Pnkm{k,\,l}{k\pm t}{l\pm t},\quad
[\mathbb{N}_2,\Pnkm{k,\,l}{k\pm t}{l\pm t}]=\pm t \Pnkm{k,\,l}{k\pm t}{l\pm t},\quad
[\mathbb{N}_1,\mathbb{N}_2]=0.
\end{equation}
Checking that the Jacobi identity true, we prove the following lemma:
\begin{lemma}
The algebras $\AGte$ and $\AGto$ are (infinite-dimensional) Lie algebras.
\end{lemma}

Using the decomposition \reff{13-1.2}, we introduce algebra $\AGt$
\begin{equation*}
\AGt=\AGte\bigoplus\AGto.
\end{equation*}
We define the Lie algebra $\AGh$ as an algebra obtained from $\AGt$ by adding of all
formal series in generators \reff{13-4.3}, including operators \reff{13-4.1},
\reff{13-4.2}, and also the operators
\begin{equation}\label{13-4.7}
\Pnkm{k}{\pm}{l}=\bigoplus_{m\geq0}\Pnkm{m\pm k,\,k}{m\pm l}{l},\qquad
\Pnkm{k}{l}{\pm}=\bigoplus_{m\geq0}\Pnkm{k,\,m\pm k}{l}{m\pm l}.
\end{equation}
The commutation relations for the algebra  $\AGh$ are induced  by the relations  \reff{13-4.4}-\reff{13-4.6}. From  \reff{13-3.3} и \reff{13-4.7} it follows that
$$
\Pnkm{k}{-}{l}=P^{(l)}_{\bullet,\,k},\quad \Pnkm{k}{l}{-}=P^{(l)}_{k,\,\bullet}.
$$
The following commutation relations
\begin{equation}\label{13-4.8}
[\mathbb{N}_1,\mathbb{N}_2]=0,\quad
[\mathbb{N}_j,\Pnkm{k,\,l}{k\pm t}{l\pm t}]=\pm t\Pnkm{k,\,l}{k\pm t}{l\pm t},
\quad [\mathbb{N}_j,\arpm]=\pm\arpm, \quad j=1,2;
\end{equation}
\begin{gather}
[\mathbb{N}_1,\,\Pnkm{k}{+}{l}]=(l-k)\Pnkm{k}{+}{l},\quad
[\mathbb{N}_2,\,\Pnkm{k}{l}{+}]=(l-k)\Pnkm{k}{l}{+},\nn\\
[\mathbb{N}_1,\,\Pnkm{k}{l}{+}]=(l-k)\Pnkm{l}{k}{+},\quad
[\mathbb{N}_2,\,\Pnkm{k}{+}{l}]=(l-k)\Pnkm{k}{+}{l};\label{13-4.9}
\end{gather}
\begin{equation}\label{13-4.10}
[\Pnkm{k,\,l}{k\pm t}{l\pm t},\Pnkm{m,\,n}{m\pm s}{n\pm s}]=
\delta_{k,\,m\pm s}\delta_{l,\,n\pm s}\Pnkm{m,\,n}{k\pm t}{l\pm t}-
\delta_{m,\,k\pm t}\delta_{n,\,l\pm t}\Pnkm{k,\,l}{m\pm s}{n\pm s};
\end{equation}

\begin{equation}\label{13-4.11}
[\Pnkm{k,\,l}{k\pm t}{l\pm t},\arpm]=\sqrt{\frac{2}{5}}
\left(\Pnkm{k\mp 1,\,l\mp 1}{k\pm t}{l\pm t}-
\Pnkm{k,\,l}{k\pm t\pm 1}{l\pm t\pm 1}\right);
\end{equation}
\begin{equation}\label{13-4.12}
[\Pnkm{k,\,l}{k\pm t}{l\pm t},\Pnkm{m}{+}{n}]=
\Pnkm{k-n+m,\,m}{k\pm t}{n\pm t}-\Pnkm{k,\,m\mp t}{k\pm t-m+n}{n};
\end{equation}
\begin{equation}\label{13-4.13}
[\Pnkm{k,\,l}{k\pm t}{l\pm t},\Pnkm{m}{n}{+}]=
\Pnkm{m,\,l-n+m}{k\pm t}{l\pm t}-\Pnkm{m\mp t,\,l}{n}{l\pm t-m+n};
\end{equation}
\begin{equation}\label{13-4.14}
[\arm,\arp]=\frac{2}{5}\left(\Pnck{0}{0}-\Pnkc{0}{0}-\Pnkm{0,\,0}{0}{0}\right);
\end{equation}
\begin{equation}\label{13-4.15}
[\arpm,\Pnkm{m}{+}{n}]=\sqrt{\frac{2}{5}}\left(\Pnkm{m}{+}{n\pm 1}-\Pnkm{m\mp 1}{+}{n}\right);
\end{equation}
\begin{equation}\label{13-4.16}
[\arpm,\Pnkm{m}{n}{+}]=\sqrt{\frac{2}{5}}\left(\Pnkm{m}{n\pm 1}{+}-\Pnkm{m\mp 1}{n}{+}\right);
\end{equation}
\begin{equation}\label{13-4.17}
[\Pnkm{k}{+}{l},\Pnkm{m}{+}{n}]=\Pnkm{m}{+}{l}-\Pnkm{k}{+}{n}
\end{equation}
\begin{equation}\label{13-4.18}
[\Pnkm{k}{+}{l},\Pnkm{m}{n}{+}]=\Pnkm{m,\,k-n+m}{n-k+l}{l}-\Pnkm{m-l+k,\,k}{n}{l-m+n};
\end{equation}
\begin{equation}\label{13-4.19}
[\Pnkm{k}{l}{+},\Pnkm{m}{n}{+}]=\Pnkm{m}{l}{+}-\Pnkm{k}{n}{+}.
\end{equation}
generate the ideal $I_r$ of the algebra  $\AGh$.
Finally, we determine the Lie algebra $\mathfrak{A}_r$ of radial oscillator as factor algebra
\begin{equation*}
\mathfrak{A}_r=\AGh/I_r.
\end{equation*}
Algebra $\mathfrak{A}_r$ is the Lie algebra with generators  \reff{13-4.1}-\reff{13-4.3}, \reff{13-4.7}
satisfying commutation relations \reff{13-4.8}-\reff{13-4.19}.
The algebra $\mathfrak{A}_r$ is a closure of the associative algebra generated by the operators \reff{13-2.29} which satisfy the commutation relations \reff{13-2.30}.

\section{Algebra of generalized boundary oscillator}

In this section we  mainly use the same scheme of reasoning as in the previous two sections.
 Therefore, we give here only a sketch of our construction.
We define the infinite-dimensional Lie algebra $\AG_0$ as
the minimal extension of the associative algebra determined by the generators \reff{13-2.36}
with the relations \reff{13-2.37}-\reff{13-2.42}. Thus $\AG_0$
is the closure of the boundary oscillator algebra.

We introduce the algebra $\widetilde{\AG}_0$, as
an associative algebra which generators \reff{13-2.36} satisfy the commutation
relations \reff{13-2.37} - \reff{13-2.42}.
Then we define infinite-dimensional Lie algebra $\widehat{\AG}_0$ as algebra
obtained from $\widetilde{\AG}_0$ by addition of all formal series in
generators \reff{13-2.36}, including operators \reff{13-2.32}.
The commutation relations of the algebra $\widehat{\AG}_0$ are induced by the relations
\reff{13-2.37} - \reff{13-2.42}.
The relations \reff{13-2.37} - \reff{13-2.42}, as well as
\begin{gather}
[\mathbb{N}_1,\acopm]=\pm \acopm,\quad [\mathbb{N}_2,\acopm]=0,\quad
[\acop ,\acom]=-\Pnkm{0,\,0}{0}{0},\nn\\
[\Pnkm{k,\,0}{m}{0},\acopm]=\Pnkm{k\mp 1,\,0}{m}{0}-\Pnkm{k,\,0}{m\pm 1}{0},\quad
[\Pnkm{0,\,k}{0}{m},\acopm]=\delta_{k,\,0}\Pnkm{\mp 1,\,0}{0}{m}-
\delta_{m,\,0}\Pnkm{0,\,k}{\pm 1}{0},
\label{13-5.1}\\
[\Pnkm{k,\,0}{0}{m},\acopm]=\Pnkm{k\mp 1,\,0}{0}{m}-\delta_{m,\,0}\Pnkm{k,\,0}{\pm 1}{0},\quad
[\Pnkm{0,\,l}{n}{0},\acopm]=\delta_{l,\,0}\Pnkm{\mp 1,\,0}{n}{0}-\Pnkm{0,\,l}{n\pm 1}{0},\nn
\end{gather}
\begin{gather}
[\mathbb{N}_1,\aocpm]=0,\quad [\mathbb{N}_2,\aocpm]=\pm \aocpm,\quad
[\aocp ,\aocm]=\Pnkm{0,\,0}{0}{0},\nn\\
[\Pnkm{k,\,0}{m}{0},\aocpm]=\delta_{k,\,0}\Pnkm{0,\,\mp 1}{m}{0}
-\delta_{m,\,0}\Pnkm{k,\,0}{0}{\pm 1},\quad
[\Pnkm{0,\,k}{0}{m},\aocpm]=\Pnkm{0,\,k\mp 1,}{0}{m}-\Pnkm{0,\,k}{0}{m\pm 1},
\label{13-5.2}\\
[\Pnkm{k,\,0}{0}{m},\aocpm]=\delta_{k,\,0}\Pnkm{0,\,\mp 1}{0}{m}-\Pnkm{k,\,0}{0}{m\pm 1},\quad
[\Pnkm{0,\,l}{n}{0},\aocpm]=\Pnkm{0,\,l\mp 1}{n}{0}-\delta_{n,\,0}\Pnkm{0,\,l}{0}{\pm 1},\nn
\end{gather}
\begin{equation}\label{13-5.3}
[\acopm,\aocp]= \Pnkm{0,\,1}{\mp 1}{0},\qquad  [\acopm,\aocm]=-\Pnkm{\pm 1,\,0}{0}{1}.
\end{equation}
generate  the ideal  $I_0$ of the algebra $\widehat{\AG}_0$.
Finally, we define the Lie algebra $\AG_0$  as the factor algebra
\begin{equation*}
\AG_0 = \widehat{\AG}_0 / I_0.
\end{equation*}
Thus algebra $\AG_0$ is the Lie algebra generated operators \reff{13-2.32}-\reff{13-2.36}
satisfying the commutation relations \reff{13-2.37}-\reff{13-2.42}, \reff{13-5.1}-\reff{13-5.3}.
This algebra is the closure of the associative algebra with generators \reff{13-2.36}
satisfying  the commutation relations \reff{13-2.37}-\reff{13-2.42}.

\section{Algebra of generalized  Chebyshev - Koornwinder oscillator}

In this section we construct and investigate the main object of our work --- the  algebra  $\AG$ of two-dimensional generalized Chebyshev-Koornwinder oscillator that is a union of sectorial, radial, and boundary oscillators. In the work  \cite{N46}, the algebra  $\AG$ was defined as a closure of the
associative algebra $\widetilde{\AG}$  generated by operators
\begin{equation}\label{13-6.1}
\mathbb{I},\, \mathbb{N}_1,\, \mathbb{N}_2,\, \aspm,\, \arpm,\, \acopm,\, \aocpm,\,
\Pnkm{k}{\pm}{l}, \Pnkm{k}{l}{\pm},\, \Pnkm{k,\,l}{m}{n}.
\end{equation}
It is assumed that indexes $k,\,l,\,m,\,n\geq0$ are fulfilled one of the following conditions
\begin{align}
1)\qquad & m+n=k+l,\quad m-n=k-l\pm 2t;\nn\\
2)\qquad & m-n=k-l,\quad m+n=k+l\pm 2t; \label{13-6.2}\\
3)\qquad & kl=0,\quad mn=0;\nn
\end{align}
and  the generators \reff{13-6.1}
satisfy the commutation relations  \reff{13-3.7} - \reff{13-3.24}, \reff{13-4.8} - \reff{13-4.19},
\reff{13-2.37} - \reff{13-2.42}, \reff{13-5.1} - \reff{13-5.3}.

We determine the algebra $\AG$ of generalized  Chebyshev - Koornwinder oscillator as infinite-dimensional Lie algebra which is a  minimal extension of the algebra $\widetilde{\AG}$.
To define the algebra $\AG$ we  consider first the algebra $\widehat{\AG}$
obtained from $\widetilde{\AG}$ by addition of all formal series in generators \reff{13-6.1}.
The commutation relations of the algebra $\widehat{\AG}$ induced by the relations \reff{13-3.7} - \reff{13-3.24}, \reff{13-4.8} - \reff{13-4.19},
\reff{13-2.37} - \reff{13-2.42}, \reff{13-5.1} - \reff{13-5.3}.
It is helpful to introduce the following notation for some of such series,
\begin{equation}\label{13-6.3}
\widehat{P}^{[k],\,l}_{[m],\,n}=\bigoplus_s  P^{s+k,\,l}_{s+m,\,n},\qquad
\widehat{P}^{k,\,[l]}_{m,\,[n]}=\bigoplus_s P^{k,\,s+l}_{m,\,s+n}.
\end{equation}

{\bf Remarks} 1) Recall that  $\Pnkm{m,\,n}{k}{l}=0$, if at least one of the
indices is negative;

2)Operators \reff{13-3.3}, \reff{13-4.7}, \reff{13-2.32} can be expressed by the operators \reff{13-6.3} as follows:
\begin{align*}
\Pnck{n}{k}&=\hPnkm{[n],\,k}{[k]}{n};
\quad\Pnkc{n}{k}=\hPnkm{k,\,[n]}{n}{[k]};
\\[9pt]
\Pnkm{k}{+}{l}& =\hPnkm{[k],\,k}{[l]}{l};
\quad\Pnkm{k}{-}{l}=\Pnck{(k)}{l}=\hPnkm{[k],\,l}{[l]}{k};
\\[9pt]
\Pnkm{k}{l}{+}& =\hPnkm{k,\,[k]}{l}{[l]};
\quad\Pnkm{k}{l}{-}=\hPnkm{l,\,[k]}{k}{[l]};
\quad\acopm  =\hPnkm{[0],\,0}{[\pm 1]}{0};\quad
\aocpm =\hPnkm{0,\,[0]}{0}{[\pm 1]}.
\end{align*}

\bigskip

The following commutation relations
\begin{equation}\label{13-6.12}
\left\{
\begin{aligned}
& [\mathbb{N}_1,\mathbb{N}_2]=0,\quad [\mathbb{N}_1,\aspm]=\mp\aspm,\quad [\mathbb{N}_1,\arm]=\pm\arpm;\\[3pt]
& [\mathbb{N}_1,\widehat{P}^{[k],\,l}_{[m],\,n}]=(m-k)\widehat{P}^{[k],\,l}_{[m],\,n};\\[5pt]
& [\mathbb{N}_1,\widehat{P}^{k,\,[l]}_{m,\,[n]}]=(m-k)\widehat{P}^{k,\,[l]}_{m,\,[n]};\\[3pt]
& [\mathbb{N}_1,\Pnkm{k,\,l}{n}{m}]=(n-k)\Pnkm{k,\,l}{n}{m};
\end{aligned}
\right.
\end{equation}
\begin{equation}\label{13-6.13}
\left\{
\begin{aligned}
&  [\mathbb{N}_2,\aspm]=\pm\aspm,\quad [\mathbb{N}_2,\arpm]=\pm\arpm,\quad
[\mathbb{N}_2,\Pnkm{k,\,l}{n}{m}]=(m-l)\Pnkm{k,\,l}{n}{m};\\[3pt]
& \left[\mathbb{N}_2,\widehat{P}^{[k],\,l}_{[m],\,n}\right]=(n-l)
\widehat{P}^{[k],\,l}_{[m],\,n};\\[5pt]
& \left[\mathbb{N}_2,\widehat{P}^{k,\,[l]}_{m,\,[n]}\right]=(n-l)
\widehat{P}^{k,\,[l]}_{m,\,[n]};
\end{aligned}
\right.
\end{equation}
\bigskip

\begin{equation}\label{13-6.14}
\left\{
\begin{aligned}
&\left[\asm,\asp\right]=
\frac{1}{10}\left(\hPnkm{[0],\,0}{[0]}{0}-\hPnkm{0,\,[0]}{0}{[0]}\right),\\[5pt]
&\left[\aspm,\arpm\right]=\pm\frac{1}{5}\hPnkm{0,\,[\mp 1]}{0}{[\pm 1]},\\[5pt]
&\left[\asmp,\arpm\right]=\pm\frac{1}{5}\hPnkm{[\mp 1],\,0}{[\pm 1]}{0},\\[5pt]
&\left[\aspm,\widehat{P}^{[k],\,l}_{[m],\,n}\right]=\frac{1}{\sqrt{10}}
\left(\widehat{P}^{[k],\,l}_{[m\mp 1],\,n\pm 1} -
\widehat{P}^{[k\pm 1],\,l\mp 1}_{[m],\,n}\right),\\[5pt]
& \left[\aspm,\widehat{P}^{k,\,[l]}_{m,\,[n]}\right]=\frac{1}{\sqrt{10}}
\left(\widehat{P}^{k,\,[l]}_{m\mp 1,\,[n\pm 1]}-
\widehat{P}^{k\pm 1,\,[l\mp 1]}_{m,\,[n]}\right),\\[5pt]
& \left[\asmp,\Pnkm{k,\,l}{m}{n}\right]=\frac{1}{\sqrt{10}}
\left(\Pnkm{k,\,l}{m\pm 1}{n\mp 1}- \Pnkm{k\mp 1,\,l\pm 1}{m}{n}\right).
\end{aligned}
\right.
\end{equation}
\bigskip

\begin{equation}\label{13-6.15}
\left\{
\begin{aligned}
& [\arm,\arp]=\frac{2}{5}\left(\hPnkm{[0],\,0}{[0]}{0}-
\hPnkm{0,\,[0]}{0}{[0]}-P^{0,\,0}_{0,\,0}\right);\\[5pt]
& \left[\arpm,\widehat{P}^{[k],\,l}_{[m],\,n}\right]=\sqrt{\frac{2}{5}}
\left(\widehat{P}^{[k],\,l}_{[m\pm 1],\,n\pm 1}-
\widehat{P}^{[k\mp 1],\,l\mp 1}_{[m],\,n}\right)\\[5pt]
& \left[\arpm,\widehat{P}^{k,\,[l]}_{m,\,[n]}\right]=\sqrt{\frac{2}{5}}
\left(\widehat{P}^{k,\,[l]}_{m\pm 1,\,[n\pm 1]}-
\widehat{P}^{k\mp 1,\,[l\mp 1]}_{m,\,[n]}\right)\\[5pt]
& \left[\armp,\Pnkm{k,\,l}{m}{n}\right]=\sqrt{\frac{2}{5}}\left(
\Pnkm{k,\,l}{m\mp 1}{n\mp 1}- \Pnkm{k\pm 1,\,l\pm 1}{m}{n}\right).
\end{aligned}
\right.
\end{equation}
\bigskip

\begin{equation}\label{13-6.16}
\left\{
\begin{aligned}
& \left[\widehat{P}^{[k],\,l}_{[m],\,n},\widehat{P}^{[q],\,s}_{[u],\,v}\right]=
\left(\widehat{P}^{[u+v-k-l-q],\,s}_{[m],\,n}
-\widehat{P}^{[m+n-q-s-k],\,l}_{[u],\,v}\right)\\[5pt]
& \left[\widehat{P}^{[k],\,l}_{[m],\,n},\widehat{P}^{q,\,[s]}_{u,\,[v]}\right]=
\left(P^{q,\,l-v+s}_{u-k+m,\,n}-P^{q-m+k,\,l}_{u,\,n-s+v}\right)\\[5pt]
& \left[\widehat{P}^{k,\,[l]}_{m,\,[n]},\widehat{P}^{q,\,[s]}_{u,\,[v]}\right]=
\left(\widehat{P}^{q,\,[-s-k-l+u+v]}_{m,\,[n]}-
\widehat{P}^{k,\,[-l+m+n-q-s]}_{u,\,[v]}\right)
\end{aligned}
\right.
\end{equation}

\begin{equation}\label{13-6.17}
\left\{
\begin{aligned}
& \left[\widehat{P}^{[k],\,l}_{[m],\,n},P^{q,\,s}_{u,\,v}\right]=
\left(P^{q,\,s}_{u+v+m-k-l,\,n}-P^{q+s+k-m-n,\,l}_{u,\,v}\right);\\[5pt]
& \left[\widehat{P}^{k,\,[l]}_{m,\,[n]},P^{q,\,s}_{u,\,v}\right]=
\left(P^{q,\,s}_{m,\,u+v+n-k-l}-P^{k,\,q+s+l-m-n}_{u,\,v}\right).
\end{aligned}
\right.
\end{equation}

\begin{equation}\label{13-6.18}
\left[\Pnkm{k,\,l}{m}{n},\Pnkm{s,\,t}{u}{v}\right]=\delta_{k,\,u}\delta_{l,\,v}\Pnkm{s,\,t}{m}{n}-
\delta_{m,\,s}\delta_{n,\,t}\Pnkm{k,\,l}{u}{v}.
\end{equation}
generate the ideal $I$ of the algebra $\widehat{\AG}$.

We define algebra $\AG$ as the factor algebra
\begin{equation*}
\AG=\widehat{\AG}/I.
\end{equation*}
The algebra $\AG$ is
the infinite-dimensional associative algebra generated by the operators
\begin{equation}\label{13-6.20}
\mathbb{I},\, \mathbb{N}_1,\, \mathbb{N}_2,\, \aspm,\, \arpm,\,
\widehat{P}^{k,\,[l]}_{m,\,[n]},\, \widehat{P}^{[k],\,l}_{[m],\,n},\, \Pnkm{k,\,l}{m}{n},
\end{equation}
satisfying commutation relations
 \reff{13-6.12}-\reff{13-6.18}.
(Recall that it is assumed that the indexes $k,\,l,\,m,\,n\geq0$, and if at
least one of the indices of operators $P$ and $\widehat{P}$ is negative,
then the corresponding operator is equal to zero).

The algebras $\widehat{\AG}$ and $\AG$ are Lie algebras because  one can check that for generators  \reff{13-6.20} are fulfilled Jacobi identities.

\section{Investigation of algebra $\AG$}
\subsection{Commutation relations between generators of the algebra  $\AG$}

For the convenience of further considerations we shall divide the set of generators \reff{13-6.20} of the algebra  $\AG$ on 4 subset.

\bigskip

\begin{equation*}%\label{13-7.1}
\text{
\begin{tabular}{|c|c|c|}
\hline
\text{type}&\text{generators}&\text{symbolic notation}\\
\hline
$I$& \rule[-7pt]{0pt}{25pt}$\mathbb{I},\, \mathbb{N}_1,\, \mathbb{N}_2$& $A_I=A$ \\
\hline
$II$& \rule[-7pt]{0pt}{25pt}$\aspm,\,\arpm$ & $A_{II}=B$\\
\hline
$III_{left}$& \rule[-7pt]{0pt}{25pt}$\hPnkm{[k],\,l}{[m]}{n}$ & $A_{III}^{left}=C^{l}$\\
\hline
$III_{right}$& \rule[-7pt]{0pt}{25pt}$\hPnkm{k,\,[l]}{m}{[n]}$  &$A_{III}^{right}=C^{r}$ \\
\hline
$IV$& \rule[-7pt]{0pt}{25pt}$\Pnkm{k,\,l}{m}{n}$&$A_{IV}=D$ \\
\hline
\end{tabular}
}
\end{equation*}
\bigskip

From relations \reff{13-6.12} - \reff{13-6.18} it follows that
\begin{equation}\label{13-7.2}
[A_1,A_2]=0,\quad [A,B_1]=B_2,\quad [A,C^l_1]=C^l_2,\quad [A,C^r_1]=C^r_2,\quad [A,D_1]=D_2;
\end{equation}
\begin{equation}\label{13-7.3}
[B_1,B_2]=C^l+C^r+D,\quad [B,C^l_1]=C^l_2+C^l_3,\quad [B,C^r_1]=C^r_2+C^r_3,
\quad [B,D_1]=D_2+D_3;
\end{equation}
\begin{equation}\label{13-7.4}
\left\{
\begin{aligned}
&[C^l_1,C^l_2]=C^l_3+C^l_4,\quad [C^r_1,C^r_2]=C^r_3+C^r_4,\quad [C^l,C^r]=D_1+D_2,\\
&\qquad\qquad\qquad\qquad[C^l,D_1]=D_2+D_3,\quad [C^r,D_1]=D_2+D_3;
\end{aligned}
\right.
\end{equation}
\begin{equation}\label{13-7.5}
[D_1,D_2]=D_3+D_4.
\end{equation}

\subsection{Center $\mathfrak{Z}(\mathfrak{A})$ of the algebra $\AG$}
An element $z$  belongs to the center $\mathfrak{Z}(\mathfrak{A})$
of the algebra $\AG$ if and only if this element commutes with all generators $X$ \reff{13-6.20} of the algebra $\AG$:
\begin{equation}\label{13-7.6}
    [z,X]=0.
\end{equation}
The element X can be represented in the following form
\begin{multline}\label{13-7.7}
    z=d_0I+d_1\mathbb{N}_1+d_2\mathbb{N}_2+d^+_s\asp+d^-_s\asm+d^+_r\arp+d^-_r\arm+\\[3pt]
    \sum_{k,\,l,\,m,\,n\geq 0}\left(d^{(L)}_{k,\,l;\,m,\,n}\widehat{P}^{[k],\,l}_{[m],\,n}+
    d^{(R)}_{k,\,l;\,m,\,n}\widehat{P}^{k,\,[l]}_{m,\,[n]}\right)+
    \sum_{k,\,l,\,m,\,n\geq 0}d^{(D)}_{k,\,l;\,m,\,n}P^{k,\,l}_{m,\,n}.\qquad
\end{multline}

Choose $X=P^{s,\,t}_{s,\,t}$ in \reff{13-7.6}.
From the commutation relations  \reff{13-6.12}-\reff{13-6.18} we get
\begin{subequations}
\begin{eqnarray}
[I, P^{s,\,t}_{s,\,t}]=0,\quad [\mathbb{N}_1, P^{s,\,t}_{s,\,t}]=0,\quad
[\mathbb{N}_2, P^{s,\,t}_{s,\,t}]=0;\qquad\qquad\qquad\qquad\qquad \label{13-7.8a} \\
\left[ \aspm , P^{s,\,t}_{s,\,t} \right]=
\frac{1}{\sqrt{10}}\left(P^{s,\,t}_{s\mp 1,\,t\pm 1}-P^{s\pm 1,\,t\mp 1}_{s,\,t}\right)\neq 0;
\qquad\qquad\qquad\qquad\qquad \label{13-7.8b}\\
\left[ \arpm , P^{s,\,t}_{s,\,t} \right]=
\sqrt{\frac{2}{5}}\left(P^{s,\,t}_{s\pm 1,\,t\pm 1}-P^{s\mp 1,\,t\mp 1}_{s,\,t}\right)\neq 0;
\qquad\qquad\qquad\qquad\qquad \label{13-7.8c}\\
\left[\widehat{P}^{[k],\,l}_{[m],\,n},\,P^{s,\,t}_{s,\,t} \right]=
\left(P^{s,\,t}_{s+t+m-k-l,\,n}-P^{s+t-m+k-n,\,l}_{s,\,t} \right)=
\left\{
\begin{aligned}
=0,&\quad\text{as}\quad m=k,\,\,n=l=t\\
\neq 0,&\quad\text{otherwise};\\
\end{aligned}
\right.
\label{13-7.8d}\\
\left[\widehat{P}^{k,\,[l]}_{m,\,[n]},\,P^{s,\,t}_{s,\,t} \right]=
\left(P^{s,\,t}_{m,\,s+t+n-k-l}-P^{k,\,s+t-m+l-n}_{s,\,t} \right)=
\left\{
\begin{aligned}
=0,&\quad\text{as}\quad n=l,\,\,m=k=s\\
\neq 0,&\quad\text{otherwise};\\
\end{aligned}
\right.
\label{13-7.8e}\\
\left[P^{k,\,l}_{m,\,n},\,P^{s,\,t}_{s,\,t} \right]=
\left\{
\begin{aligned}
P^{s,\,t}_{m,\,n},&\quad\text{as}\quad (k,l)=(s,t)\neq(m,n)\\
-P^{k,\,l}_{s,\,t},&\quad\text{as}\quad (m,n)=(s,t)\neq(k,l)\\
 0,&\quad\text{otherwise};\\
\end{aligned}
\right.
\qquad\qquad\qquad\label{13-7.8f}
\end{eqnarray}
\end{subequations}
We remind that all generators \reff{13-6.20}  are linearly independent. Then, from \reff{13-7.6}  it follows that all coefficients at nonzero commutators must be equal to zero.
This allows us to simplify the \reff{13-7.7} as follows
\begin{multline}\label{13-7.9}
   \qquad\qquad z=d_0I+d_1\mathbb{N}_1+d_2\mathbb{N}_2+
    \sum_{k\geq 0}
    d^{(L)}_{k,\,t;\,k,\,t}\widehat{P}^{[k],\,t}_{[k],\,t}+
    \sum_{l\geq 0}
    d^{(R)}_{s,\,l;\,s,\,l}\widehat{P}^{s,\,[l]}_{s,\,[l]}
    +\\
    \sum_{{\begin{matrix}(k,l)\neq(s,t)\neq(m,n)\\(k,l)\neq(m,n)\end{matrix}}}
   \hspace{-1cm} d^{(D)}_{k,\,l;\,m,\,n}P^{k,\,l}_{m,\,n}+
    \sum_{k,\,l\geq 0}d^{(D)}_{k,\,l;\,k,\,l}P^{k,\,l}_{k,\,l}.\qquad\qquad\qquad\qquad
\end{multline}
We can put $(s,\,t)=(k,\,l)$ or $(s,\,t)=(m,\,n)$, because  $s,\,t$ are arbitrary integer numbers.
Then we can prove that all coefficients $d^{(D)}_{k,\,l;\,m,\,n}$ in
the penultimate sum in the right-hand side of \reff{13-7.9} equals to zero.
Thus \reff{13-7.9} become simpler and  takes the form
\begin{equation}\label{13-7.10}
z=d_0I+d_1\mathbb{N}_1+d_2\mathbb{N}_2
+\sum_{k\geq 0}d^{(L)}_{k,\,t;\,k,\,t}\widehat{P}^{[k],\,t}_{[k],\,t}
+\sum_{l\geq 0}d^{(R)}_{s,\,l;\,s,\,l}\widehat{P}^{s,\,[l]}_{s,\,[l]}
+\sum_{k,\,l\geq 0}d^{(D)}_{k,\,l;\,k,\,l}P^{k,\,l}_{k,\,l}.
\end{equation}
Let X be one of ladder operators, for example $\asp$.
Then, from \reff{13-7.8b}, \reff{13-6.14} and
\begin{equation*}
[\mathbb{N}_1,\asp]=-\asp\neq 0,\quad    [\mathbb{N}_2,\asp]=\asp\neq 0,
\end{equation*}
we conclude that all coefficients in the right-hand side of the equality \reff{13-7.10},
excluding  $d_0$, are equal to zero.
So, we have $z=d_0 I$ that is,
\begin{equation*}
\mathfrak{Z}(\mathfrak{A})=d_0 I,
\end{equation*}
which means that the center $\mathfrak{Z}(\mathfrak{A})$
of the algebra $\mathfrak{A}$ is one-dimensional.

\subsection{ Maximal Abelian subalgebra of the algebra $\AG$}
We consider the commutative subalgebra  $\mathfrak{L}$
in $\mathfrak{A}$, generated by the operators
 $\left\{ P^{m,\,n}_{m,\,n}\right\}_{m,\,n\geq 0}$.
We denote by $\mathfrak{M}$ Abelian subalgebra in  $\mathfrak{A}$,
obtained from $\mathfrak{L}$ by addition to generators $P^{m,\,n}_{m,\,n}$
of all formal series in these generators. Note that
\begin{equation}\label{13-7.12}
\mathbb{N}_1=\bigoplus_{m,\,n\geq0} m P^{m,\,n}_{m,\,n},\quad
\mathbb{N}_2=\bigoplus_{m,\,n\geq0} n P^{m,\,n}_{m,\,n},\quad
I=\bigoplus_{m,\,n\geq0}  P^{m,\,n}_{m,\,n},
\end{equation}
which means that an arbitrary element $u\in\mathfrak{M}$ can be written as
\begin{equation}\label{13-7.13}
u=\bigoplus_{k,\,l\geq 0}c_{k,\,l}P^{k,\,l}_{k,\,l}.
\end{equation}
We show that  $\mathfrak{M}$ is a maximal Abelian subalgebra in Lie algebra $\mathfrak{A}$.
Indeed, suppose that there is some element  $z$  of $\mathfrak{A}$  such that
\begin{equation*}
[z,X]=0,
\end{equation*}
for all $X\in \mathfrak{M}$.
Choosing $X=P^{s,\,t}_{s,\,t}$
(see reasons above),
we obtain for $z$  the equality \reff{13-7.10}, i.e.
\begin{equation}\label{13-7.15}
z=d_0I+d_1\mathbb{N}_1+d_2\mathbb{N}_2
+\sum_{k\geq 0}d^{(L)}_{k,\,t;\,k,\,t}\widehat{P}^{[k],\,t}_{[k],\,t}
+\sum_{l\geq 0}d^{(R)}_{s,\,l;\,s,\,l}\widehat{P}^{s,\,[l]}_{s,\,[l]}
+\sum_{k,\,l\geq 0}d^{(D)}_{k,\,l;\,k,\,l}P^{k,\,l}_{k,\,l}.
\end{equation}
Taking into account \reff{13-6.3} and \reff{13-7.12}, the relation \reff{13-7.15} can be rewritten as \begin{equation*}
z=\bigoplus_{k,\,l\geq0}\left(k+l+d^{(D)}_{k,\,l;\,k,\,l}\right)P^{k,\,l}_{k,\,l}
\bigoplus_{k\geq0}\left(\sum_{s=0}^k d^{(L)}_{s,\,t;\,s,\,t}\right)P^{k,\,t}_{k,\,t}
\bigoplus_{l\geq0}\left(\sum_{t=0}^l d^{(R)}_{s,\,t;\,s,\,t}\right)P^{s,\,l}_{s,\,l}.
\end{equation*}
In view of \reff{13-7.13},  this means that $z\in\mathfrak{M}$.
Hence it follows that subalgebra  $\mathfrak{M}$ is the maximal Abelian subalgebra in the Lie algebra $\mathfrak{A}$.

Note that $\mathfrak{M}$ contains the commutative subalgebra constructed in work
 \cite{N47}. This subalgebra is an extension of Koornwinder algebra \cite{N28}
consisting of all differential operators of the variables  $z,\,\overline{z}$,
for which Chebyshev-Koornwinder polynomials  are eigenfunctions.

\subsection{Subalgebras and ideals in $\AG$. Representation $\AG$ in the form  of semidirect sum}
Let us consider the following vector subspace of the algebra $\mathfrak{A}$:
\begin{align*}
\mathfrak{A_1}&=\overline{\text{Span}}\{\mathbb{I}, \mathbb{N}_1, \mathbb{N}_2\};
\\
\mathfrak{A_2}&=\overline{\text{Span}}\{\aspm, \arpm\};
\\
\mathfrak{A}_3^L&=\overline{\text{Span}}\{\widehat{P}^{k,\,[l]}_{m,\,[n]}\},\qquad
\mathfrak{A}_3^R=\overline{\text{Span}}\{\widehat{P}^{[k],\,l}_{[m],\,n}\},\quad k,l,m,n\geq 0;
\\
\mathfrak{A}_3&=\mathfrak{A}_3^L\boldsymbol{\dot{+}}\mathfrak{A}_3^R,
\\
\mathfrak{A}_4&=\overline{\text{Span}}\{P^{k,\,l}_{m,\,n}\},
\quad k,l,m,n\geq 0;
\end{align*}
\begin{equation*}
\mathfrak{B}_1^L=\mathfrak{A_2}\boldsymbol{\dot{+}}\mathfrak{A}_3^L,\qquad
\mathfrak{B}_1^R=\mathfrak{A_2}\boldsymbol{\dot{+}}\mathfrak{A}_3^R.
\end{equation*}
\begin{equation*}
\mathfrak{B}_1=\mathfrak{B}_1^L\boldsymbol{\dot{+}}\mathfrak{B}_1^R,
\end{equation*}
\begin{equation*}
\mathfrak{B}_2^L=
\mathfrak{A_2}\boldsymbol{\dot{+}}\mathfrak{A}_3^L\boldsymbol{\dot{+}}\mathfrak{A}_4,\qquad
\mathfrak{B}_2^R=
\mathfrak{A_2}\boldsymbol{\dot{+}}\mathfrak{A}_3^R\boldsymbol{\dot{+}}\mathfrak{A}_4,
\end{equation*}
\begin{equation*}
\mathfrak{B}_2=\mathfrak{B}_2^L\boldsymbol{\dot{+}}\mathfrak{B}_2^R,
\end{equation*}

\begin{equation*}
\mathfrak{B}_3^L=\mathfrak{A}_3^L\boldsymbol{\dot{+}}\mathfrak{A}_4,\qquad
\mathfrak{B}_3^R=\mathfrak{A}_3^R\boldsymbol{\dot{+}}\mathfrak{A}_4,
\end{equation*}
\begin{equation*}
\mathfrak{B}_3=\mathfrak{B}_3^L\boldsymbol{\dot{+}}\mathfrak{B}_3^R,
\end{equation*}

Of the commutation relation \reff{13-7.2}-\reff{13-7.5},
it follows that
\begin{itemize}
  \item $\mathfrak{A}_1$ --- Abelian subalgebra of infinite-dimensional algebra $\mathfrak{A}$;
  \item $\mathfrak{B}_2^L$ ,\, $\mathfrak{B}_2^R$ ,\,
  $\mathfrak{B}_2^L\bigoplus\mathfrak{B}_2^R$
   --- two-sided ideals in $\mathfrak{A}$, where $\bigoplus$ ---
   direct sum of ideals;
  \item $\mathfrak{B}_3^L$  --- two-sided ideals in $\mathfrak{A}$ and therefore
  two-sided ideals in $\mathfrak{B}_2$ and in $\mathfrak{B}_2^L$;
  \item $\mathfrak{B}_3^R$  --- two-sided ideals in $\mathfrak{A}$ and therefore
  two-sided ideals in
  $\mathfrak{B}_2$ and in$\mathfrak{B}_2^R$;
  \item $\mathfrak{A}_4$  ---  two-sided ideals in $\mathfrak{A}$ and therefore
  two-sided ideals in $\mathfrak{B}_2$;
\end{itemize}
Then $\mathfrak{A}_4$ is two-sided ideals in $\mathfrak{B}_2^L$
and in $\mathfrak{B}_2^R$, as well in  $\mathfrak{B}_3$, $\mathfrak{B}_3^L$
and in $\mathfrak{B}_3^R$ too.
\bigskip
We construct derived series for the algebra  $\mathfrak{A}$
\begin{equation*}
\mathfrak{A}^{(0)}=\mathfrak{A},\quad
\mathfrak{A}^{(1)}=[\mathfrak{A}^{(0)},\mathfrak{A}^{(0)}],\quad
\mathfrak{A}^{(2)}=[\mathfrak{A}^{(1)},\mathfrak{A}^{(1)}],\quad \ldots .
\end{equation*}
It can be checked that
\begin{equation}\label{13-7.28}
\mathfrak{A}^{(1)}=\mathfrak{B}_2,\qquad
\mathfrak{A}^{(n)}=\mathfrak{A}^{(2)}=\mathfrak{B}_3,\quad \forall n\geq2.
\end{equation}
From \reff{13-7.28} it follows that $\mathfrak{A}$ is not solvable
and therefore also not nilpotent algebra.

Denote by $\mathbb{N}$ a radical (i.e. maximal solvable ideal) of algebra $\mathfrak{A}$.
It is known that
$\mathfrak{Z}(\mathfrak{A})\subseteq \mathbb{N}$. Then, taking into account \reff{13-7.28},
we have
\begin{equation*}
\mathfrak{Z}(\mathfrak{A})\subseteq \mathbb{N}\subset\mathfrak{A}.
\end{equation*}
To verify that the generalization of the theorem Levi-Maltsev \cite{N48} true for algebra $\mathfrak{A}$ , we must find the radical $\mathbb{N}$ of the algebra $\mathfrak{A}$.

{\bf Remark}. The first step in finding $\mathbb{N}$  is to construct
solvable ideal $\mathbb{L}\subseteq\mathbb{N}$ such that
\begin{equation}\label{13-7.30}
    [\mathbb{L},\mathbb{L}]=\mathfrak{Z}(\mathfrak{A})\quad\Rightarrow\quad
    \mathfrak{Z}(\mathfrak{A})\subseteq\mathbb{L}.
\end{equation}
There are two  alternatives
\begin{subequations}
\begin{eqnarray}
\mathbb{L}\quad\text{does not exist} \quad\Rightarrow\quad
\mathbb{N}=\mathfrak{Z}(\mathfrak{A}) \label{13-7.31a},\\
\mathbb{L}\quad\,\text{exist}\, \quad\Rightarrow\quad
\mathfrak{Z}(\mathfrak{A})\subset\mathbb{N}\label{13-7.31b}.
\end{eqnarray}
\end{subequations}
Our hypothesis is true \reff{13-7.31a}, i.e $\mathfrak{A}$ reductive \cite{N48a}, but at present we can not give the proof of validity of this hypothesis.

\bigskip

Recall that a Lie algebra $\mathfrak{L}$  is the semi-direct sum of Lie subalgebras  $\mathfrak{T}$ and $\mathfrak{M}$ ($\mathfrak{L}=\mathfrak{M}\sdsum \mathfrak{T}$) if
\begin{gather}
\mathfrak{L}=\mathfrak{M}\boldsymbol{\dot{+}}\mathfrak{T}, \label{13-7.32}\\
    [\mathfrak{T},\mathfrak{T}]\subset \mathfrak{T},\quad
    [\mathfrak{M},\mathfrak{M}]\subset \mathfrak{M},\quad
    [\mathfrak{M},\mathfrak{T}]\subset \mathfrak{T}. \label{13-7.33}
\end{gather}
As seen from \reff{13-7.33}, $\mathfrak{T}$ is an ideal in $\mathfrak{L}$.

Since $\mathfrak{A}=\mathfrak{A}_1\boldsymbol{\dot{+}}\mathfrak{B}_2$,
where $\mathfrak{A}_1$ is a subalgebra in  $\mathfrak{A}$
and $\mathfrak{B}_2$ is an ideal in  $\mathfrak{A}$, then from \reff{13-7.32}, \reff{13-7.33} we have
\begin{equation*}
\AG=\mathfrak{A}_1\sdsum\mathfrak{B}_2.
\end{equation*}

We introduce notations
\begin{equation*}
\mathfrak{A}^L=\mathfrak{A}_1\boldsymbol{\dot{+}}\mathfrak{B}_2^L,\qquad
\mathfrak{A}^R=\mathfrak{A}_1\boldsymbol{\dot{+}}\mathfrak{B}_2^R,
\end{equation*}
where $\mathfrak{A}_1$ is a subalgebra in $\mathfrak{A}^L$ and in $\mathfrak{A}^R$;
$\mathfrak{B}_2^L$ is an ideal in  $\mathfrak{A}^L$; $\mathfrak{B}_2^R$ is an ideal in $\mathfrak{A}^R$.
Then, from \reff{13-7.32}, \reff{13-7.33} implies that
\begin{equation*}
\mathfrak{A}^L=\mathfrak{A}_1\sdsum\mathfrak{B}_2^L,\qquad
\mathfrak{A}^R=\mathfrak{A}_1\sdsum\mathfrak{B}_2^R.
\end{equation*}
In addition, as
\begin{equation*}
\mathfrak{B}_2^L=\mathfrak{B}_1^L\boldsymbol{\dot{+}}\mathfrak{A}_4,\qquad
\mathfrak{B}_2^R=\mathfrak{B}_1^R\boldsymbol{\dot{+}}\mathfrak{A}_4,
\end{equation*}
where $\mathfrak{B}_1^L$ is a subalgebra in $\mathfrak{B}_2^L$;
$\mathfrak{B}_1^R$ is a subalgebra in $\mathfrak{B}_2^R$;
$\mathfrak{A}_4$ is an ideal in $\mathfrak{B}_2^L$ and in $\mathfrak{B}_2^R$, then
\begin{equation*}
\mathfrak{B}_2^L=\mathfrak{B}_1^L\sdsum\mathfrak{A}_4,\qquad
\mathfrak{B}_2^R=\mathfrak{B}_1^R\sdsum\mathfrak{A}_4.
\end{equation*}
However note that,  $\mathfrak{B}_2\neq\mathfrak{B}_1\sdsum\mathfrak{A}_4$,
as $\mathfrak{B}_1^L + \mathfrak{B}_1^R=\mathfrak{B}_1$
is not an subalgebra in $\mathfrak{B}_2$.

\subsection{Proof simplicity of the ideal $\mathfrak{A}_4$}

We prove that $\mathfrak{A}_4$ is the simple ideal.
To do this, we suppose that there exists a non-zero two-sided ideal
$I\subset\mathfrak{A}_4$ and show that
\begin{equation}\label{13-7.39}
    I=\mathfrak{A}_4.
\end{equation}
To begin with, we choose the standard basis in $\mathfrak{A}_4$
\begin{equation*}
\left\{ P^{k,\,l}_{m,\,n}, (k,\,l)\neq(m,\,n); h_{m,\,0},\, h_{0,\,n},\, h_{m,\,n}\right\},
\end{equation*}
where $h_{m,\,0}=P^{m,\,0}_{m,\,0}$,\, $h_{0,\,n}=P^{0,\,n}_{0,\,n}$,\,
$h_{m,\,n}=P^{m+1,\,n+1}_{m+1,\,n+1}-P^{m,\,n}_{m,\,n}$.

To prove that this set is really forms the basis of the $\mathfrak{A}_4$, we should check that any element $P^{m,\,n}_{m,\,n}$ ($m\geq0,\, n\geq0$) belongs to the linear span of this set.
But this assertion follows from the obvious equality
\begin{equation*}
 P^{m,\,n}_{m,\,n}=\sum_{k=0}^{\min(m,\,n)} h_{m-k,\, n-k}.
\end{equation*}

Further, if $I$ is ideal distinct from zero, then there exists a nonzero element $z\in I$.
We present $z$ in the form of a linear combination of the basis elements
\begin{equation}\label{13-7.41}
    z=\sum_{\begin{matrix}k,l,m,n \geq 0\\ (k,l)\neq(m,n)\end{matrix}}
    \alpha^{k,l}_{m,n}P^{k,l}_{m,n}+\sum_{m,n\geq0}\beta_{m,n}h_{m,n}.
\end{equation}
Let $X=P^{s,t}_{u,v}$, ($(u,v)\neq (s,t)$) is an  element of $\mathbb{D}$.
Then
  \begin{equation*}
    (\text{ad}_X^2)z=-2\alpha^{u,v}_{s,t}X
\end{equation*}
Since $z\in I$, it follows that $\alpha^{u,v}_{s,t}X\in I$, and if $\alpha^{u,v}_{s,t}\neq0$, then  $X=P^{s,t}_{u,v}\in I$.
Doing so for all elements $P^{s,t}_{u,v}$ with $(u,v)\neq (s,t)$,
we get two alternatives
\begin{enumerate}
  \item There is a nonzero element $P^{s,t}_{u,v}\in I$.
  \item All the coefficients $\alpha^{k,l}_{m,n}$ in the expansion \reff{13-7.41}
  are equal to zero, and then
\begin{equation*}
  z=\sum_{m,n\geq0}\beta_{m,n}h_{m,n}\in I.
  \end{equation*}
\end{enumerate}
\medskip
 We have to show that in both cases, equation  \reff{13-7.39} is true , i.e.  $I=\mathfrak{A}_4$.

\bigskip

In the first case to check the validity of equality  \reff{13-7.39} it is enough to show that
if $X=P^{s,t}_{u,v}\in I$ under the condition $(s,t)\neq(u,v)$, then any basic element
$P^{k,l}_{m,n}$ (with $(k,l)\neq(m,n)$)
and any basic element $h_{m,n}$ belong $I$.
Using commutation relations
\begin{equation*}
\left[P^{s,t}_{u,v},P^{k,l}_{s,t}\right]=P^{k,l}_{u,v}\in I,\quad (k,l)\neq(u,v);
\end{equation*}
we obtain
\begin{equation*}
\left[P^{k,l}_{u,v},P^{u,v}_{m,n}\right]= - P^{k,l}_{m,n}\in I.
\end{equation*}
If $(k,l)=(u,v)$, then we have
\begin{equation*}
\left[P^{s,t}_{k,l},P_{s,t}^{m,n}\right]=P_{k,l}^{m,n}\in I,\quad (k,l)\neq(m,n).
\end{equation*}
Further, it is straightforward to show that if $(k,l)\neq (m,n)$
from $P^{m,n}_{k,l}\in I$ implies that $P^{k,l}_{m,n}\in I$.
It remains to show that any element $h_{m.n}\in I$.
Since for  $(s,t)\neq(u,v)$,
\begin{equation*}
    P^{s,t}_{u,v}\in I\quad \Rightarrow P^{u,v}_{s,t}\in I,
\end{equation*}
then
\begin{equation}\label{13-7.45}
\left[P^{s,t}_{u,v},P^{u,v}_{s,t}\right]=\left(P^{u,v}_{u,v}-P^{s,t}_{s,t}\right)\in I.
\end{equation}
For $u=m+1,\, v=n+1,\, s=m,\, t=n$ from \reff{13-7.45} it follows that
\begin{equation}\label{13-7.46}
h_{m,n}=\left(P^{m+1,n+1}_{m+1,n+1}-P^{m,n}_{m,n}\right)\in I.
\end{equation}
Note that \reff{13-7.46} is true for both $h_{m,0}$ and $h_{0,n}$.
So, the first variant considered fully.

Let us now consider the second variant.
In this case there exists at least one non-null element $h_{m,n}\in I$.
Then if $(s,t)\neq(m,n)$ and $(s,t)\neq(m+1,n+1)$, we have for any $P^{s,t}_{m,n}$
\begin{equation*}
\left[h_{m,n},P^{s,t}_{m,n}\right]=
\left[\left(P^{m+1,n+1}_{m+1,n+1}-P^{m,n}_{m,n}\right),P^{s,t}_{m,n}\right]=
-P^{s,t}_{m,n}\in I.
\end{equation*}
Using the facts proved in considering the first variant, we get
 $I=\mathfrak{A}_4$.
So, the relation \reff{13-7.39} is proven.
Since  $\mathfrak{A}_4$
has no nonzero ideals other than $\mathfrak{A}_4$,
then  $\mathfrak{A}_4$ is simple algebra.

\section{Conclusion}

\hspace{0.5cm}1. One can  consider presented in the paper ChK-oscillator
as the simplest nontrivial example of a quantum system composed of three
interacting one-dimensional oscillators.
Note that in all known to the authors papers on generalized oscillators associated with orthogonal polynomials in several variables these oscillators form a system of independent
one-dimensional oscillators, because the related oscillator algebras splits
in direct sum of classical  Lie algebras.

2. An interesting question is under which conditions the oscillator algebra
is finite-dimensional.
 The answer to this question was given in the work \cite{N49}
for an one-dimensional generalized oscillator related to
a system of polynomials  orthogonal with respect to symmetric measure on real axis.
In the cited work were given some consideration of the oscillator algebras associated with multi-boson systems.

The results of this work suggest that in the nontrivial case
(i.e. when a multi-dimensional oscillator describes a system of interacting particles)
the corresponding oscillator algebra is infinite-dimensional.

3. In conclusion, we note that in our work was made only the first step in study
of Lie algebra of two-dimensional ChK-oscillator.
In particular, it was necessary to investigate the possibility of constructing
the root system for considered infinite-dimensional Lie algebra.
For such investigation one must to find the radical subalgebra of oscillator algebra
and to check the validity of Levi-Maltsev decomposition. The authors
intend to investigate this question in subsequent publications.

\section*{Acknowledgements}
We like to thank P.P. Kulish and V.D. Lyakhovsky for discussions and comments.
The work of EVD done under the partial support of the RFBR grant 12-01-00207а.
%\begin{acknowledgments}
%We wish to acknowledge
%\end{acknowledgments}

\end{document}